\documentclass[journal]{IEEEtran}

\ifCLASSINFOpdf
  \usepackage[pdftex]{graphicx}
  \DeclareGraphicsExtensions{.pdf,.jpeg,.png}
\else
  \usepackage[dvips]{graphicx}
\fi
\usepackage{epstopdf}
\usepackage[compress,nospace]{cite}
\usepackage{amsthm}
\usepackage{stfloats}
\usepackage[cmex10]{amsmath}
\usepackage{amsmath}
\usepackage{cases}
\usepackage{siunitx}
\usepackage{amssymb}
\usepackage{wasysym}
\usepackage{algorithm}
\usepackage{algorithmic}
\usepackage[colorlinks,linkcolor=blue,anchorcolor=blue,citecolor=blue,bookmarks=blue]{hyperref}
\usepackage{array}
\usepackage{url}
\usepackage{color}
\usepackage{subfigure}
\usepackage{verbatim}
\usepackage{multicol}  
\usepackage{multirow} 
\usepackage{booktabs}
\usepackage{makecell}
\usepackage{setspace}

\begin{document}
	
\title{Learning-Aided Beam Prediction in mmWave MU-MIMO Systems for High-Speed Railway}

\author{Fan~Meng, Shengheng~Liu,~\IEEEmembership{Member,~IEEE}, Yongming~Huang,~\IEEEmembership{Senior Member,~IEEE}, Zhaohua~Lu
	\thanks{This work was supported in part by the National Key R\&D Program of China under Grant 2020YFB1806600, the National Natural Science Foundation of China under Grants 61720106003 and 62001103, and the Research Project of Jiangsu Province under Grant BE2018121. Part of this work has been accepted for presentation at the IEEE Global Communications Conference (GLOBECOM): Signal Processing for Communications Symposium, Madrid, Spain, December 2021 \cite{GlobeCom21}. (Corresponding authors: Y.\ Huang. and Z.\ Lu.)}
	\thanks{Fan~Meng is with the Purple Mountain Laboratories, Nanjing 211111, China (e-mail: mengfan@pmlabs.com.cn)}
	\thanks{Shengheng Liu and Yongming Huang are with the National Mobile Communications Research Laboratory, School of Information Science and Engineering, Southeast University, Nanjing 210096, China, and also with the Purple Mountain Laboratories, Nanjing 211111, China (e-mail: s.liu@seu.edu.cn; huangym@seu.edu.cn).}
	\thanks{Zhaohua Lu is with the ZTE Corporation, Shenzhen 518057, China, and also with State Key Laboratory of Mobile Network and Mobile Multimedia Technology, Shenzhen 518057, China (e-mail: lu.zhaohua@zte.com.cn).}	
}

\maketitle

\begin{abstract}

The problem of beam alignment and tracking in high mobility scenarios such as high-speed railway (HSR) becomes extremely challenging, since large overhead cost and significant time delay are introduced for fast time-varying channel estimation. To tackle this challenge, we propose a learning-aided beam prediction scheme for HSR networks, which predicts the beam directions and the channel amplitudes within a period of future time with fine time granularity, using a group of observations. Concretely, we transform the problem of high-dimensional beam prediction into a two-stage task, i.e., a low-dimensional parameter estimation and a cascaded hybrid beamforming operation. In the first stage, the location and speed of a certain terminal are estimated by maximum likelihood criterion, and a data-driven data fusion module is designed to improve the final estimation accuracy and robustness. Then, the probable future beam directions and channel amplitudes are predicted, based on the HSR scenario priors including deterministic trajectory, motion model, and channel model. Furthermore, we incorporate a learnable non-linear mapping module into the overall beam prediction to allow non-linear tracks. Both of the proposed learnable modules are model-based and have a good interpretability. Compared to the existing beam management scheme, the proposed beam prediction has (near) zero overhead cost and time delay. Simulation results verify the effectiveness of the proposed scheme.

\end{abstract}


\section{Introduction}\label{sec:introduction}

In the fifth-generation (5G) and future wireless communications, millimeter wave (mmWave) communication arises as an appealing solution to provide abundant available spectrum, and thus satisfies the critical demands for the explosively growing data traffic~\cite{5490974, 7959169}. However, data transmission in the mmWave band is challenging due to the high path loss, resulting in a limited coverage area. The small carrier wavelength enables packing a large number of antenna elements into small form factors. Leveraging the large antenna arrays employed at the transmitter and receiver, mmWave systems can perform directional beamforming to achieve high beamforming gain, which helps overcome large free-space path loss of mmWave signals and guarantees sufficient received signal-to-noise ratio (SNR). Nevertheless, the large-scale antennas bring significant challenges for channel estimation, especially in highly mobile environments.

Recently, deep learning (DL)~\cite{Goodfellow2016Deep} has been applied to physical layer communications and regarded as an enabling technology for future wireless mobile network. The learning-based approach is data-driven, and inherently applicable for the scenarios with imperfect models and/or intractable problems, where the model-driven method cannot work well~\cite{8054694, 9120241, 8353153, 9018199}. However, the data-driven method especially the end-to-end scheme has several obvious shortcomings, including high dependence on data, high training and model complexity, lack of interpretability and performance guarantee. Meanwhile, model-driven methods are free from these shortcomings by nature. Therefore, embedding learnable modules into the existing model-based system, or designing a specific neural network (NN) with domain knowledge in communications can combine the advantages of both paradigms and possibly achieve better performance~\cite{8742579, 8715338}.

In terms of beam alignment/tracking (BA/T) for mobile terminals (MTs), current model-based methods are feasible, and achieve (sub)-optimal performance with simple and explicit simulated models~\cite{6717211, 7390019, 2017Location}. Meanwhile, the practical environments have implicit and complex prior information in the time frequency spatial domain, which the data-driven methods can better utilize than the model-driven ones. In the literature, the learning-enabled BA/T in mobile environments have been widely-investigated in recent years. The beam alignment and user localization are strongly coupling in mmWave communications, With the aid of spatial location information, it is possible to conduct beam alignment with higher accuracy and lower overhead. In \cite{8823977}, a mapping from the user location to the beam pairs (fingerprints) is learned by supervised learning (SL). The labeled data are collected by different locations and stored in a database, and the mapping is usually realized by a deep NN (DNN) in complex practical environments. The spatial location information also can be implicitly presented as global positioning system (GPS) signals~\cite{9013296} and 3-D point clouds~\cite{9129762}. The DL-enabled compressed sensing (CS) is developed in \cite{jonathan2019deep}, and researchers design a structured DNN-based CS matrix for vehicular environments. Except for the above SL approaches, BA/T can be realized by deep reinforcement learning (DRL)~\cite{Sutton} in a closed-loop manner~\cite{8842625, 9069211, 9269463}. In \cite{9269463}, an interactive learning design paradigm which makes full use of domain knowledge and adaptive learning, is developed. The paradigm requires no prior knowledge of the dynamic channel modeling, and thus is applicable for a variety of complicated scenarios. Different from above DL-based approaches, sparse Bayesian learning (SBL) has also been considered in \cite{8410591}, and low-rank property of time-varying massive multiple-input multiple-output (MIMO) channel covariances is utilized to reduce the training overhead. Expectation maximization-based SBL framework is used to learn the sparse parameter set. Furthermore, a Kalman filter is adopted to exploit the channel temporal correlations to enhance channel tracking accuracy. For high-speed railway (HSR) wireless networks, significant angle offset induced in initial access process is investigated in \cite{8984259}. This research is established on the periodicity and regularity of trains' trajectory. To compensate the angle offset, the aligned beam is adjusted by the historical beam training results. To reduce the beam search space, a best beam pair look-up table is learned from the historical information.

In this paper, we investigate the learning-aided BAT for HSR mmWave wireless networks. The advanced HSR system has following notable features including: high-speed MTs up to $ 500 \; \textup{km/h} $; high-density MTs up to hundreds on one carriage; and high-quality services such as real-time video transmission~\cite{7295467, 7811845}. Meanwhile, the current mobile network for the HSR system is far from satisfactory due to the scarcity of the spectrum resources. Therefore, it is essential to develop the mmWave techniques for the explosively growing demand in the advanced HSR system. The current beam management procedure which includes beam measurement, reporting and indication, performs well in a regular mmWave scenario where the MTs moves at a low speed, but is inapplicable for a typical HSR scenario~\cite{8458146}. In high-speed mobile scenarios, this procedure is inefficient due to the following two reasons:
\begin{itemize}
	\item \textbf{Beam training overhead}. Regarding beam measurement, the overhead caused by frequent beam training can be very huge, due to small beam dwelling time. When number of MTs increases to $ 50 $ and train speed is $ 500 \; \textup{km/h} $, simulation results in~\cite{beam_manage} show that almost all time frequency resource are occupied by beam training.	
	\item \textbf{Time delay loss}. Regarding beam reporting and beam indication, the corresponding latency is mainly produced by activating candidate beams from radio resource control (RRC) pool. The report~\cite{beam_manage} demonstrates that the latency can be up to $ 25 \; \textup{ms} $ with $ 20 \; \textup{ms} $ synchronization signal block (SSB) periodicity.
\end{itemize}
To the best of our knowledge, the above two problems have not been addressed in the existing studies. Therefore, to reduce the beam training overhead and time delay loss in the HSR scenarios, it is essential to develop a new beam management framework.

In this paper, we propose a learning-aided beam prediction scheme. More concretely, given a group of received pilot signals and measurements including Doppler frequencies and communication delays at different instants, we predict the optimal Tx/Rx beams within a period of future time with fine time granularity. The duration of beam prediction up to a second level, reduces the overhead and delay to be (near) zero; and the time granularity up to a millisecond level (greatly smaller than the beam dwelling time), guarantees the beamforming performance. The beam prediction can be carried out in a purely model-driven manner, but it cannot perform well with implicit environment priors and system models. On the other hand, the purely data-driven approach which outputs high-dimensional beam indexes, is difficult to be realized by an end-to-end DNN. Consequently, we innovatively propose a model-based learnable beam prediction scheme, which equivalently transform the high-dimensional beam prediction into two cascaded stages, i.e., parameter estimation and hybrid beamforming. 

First, given a group of observations, we derive estimation of two parameter sets, i.e., the MT locations and speeds separately and independently. Meanwhile, the bias and variance of the estimated results cannot be derived in a practical environment. Therefore, we propose a learnable data fusion module to implicitly estimate the corresponding bias and variance, to further improve both the estimation accuracy and robustness. Secondly, due to the prior information that the determinacy of the moving train trajectory and the mmWave channel can be well-described as urban macro (UMa) line of sight (LoS) in $ 3 \textup{GPP} $ TR $ 38.901 $~\cite{38901}, the hybrid beamforming is realized by the estimated parameter set. Additionally, to handle the non-linearity of tracks, we propose a learnable non-linear mapping module. The technical contributions of this work are summarized as follows.
\begin{itemize}
	\item We propose a beam prediction scheme which reduces the overhead and delay arised by beam measurement and reporting to (near) zero. Then, the high-dimensional beam prediction problem is equivalently transformed into two cascaded sub-problems, i.e., parameter estimation and hybrid beamforming, which are both model-based and learnable.
	\item Separate estimation of two parameter sets is performed using the maximum likelihood (ML) criterion. Furthermore, we propose a data fusion module to learn the corresponding biases and variances, and obtain a final parameter set with higher accuracy and better robustness.
	\item We propose to predict the optimal BS analog precoder and MT combiner with the estimated parameter set. The long-term prediction duration is up to $ 1.25 \; \textup{s} $, and the fine time granularity is $ 1.25 \; \textup{ms} $. The BS digital precoder is realized by classical minimal mean square error (MMSE) precoding.
	\item We propose a learnable non-linear mapping module to fit the non-linear tracks, where the approximator is composed of piece-wise linear functions. The learned mapping is used for MT location search in ML estimators, BS analog precoding and MT combining. The upper bound of fitting error is also given.
\end{itemize}
 
The rest of this paper is organized as follows. The system model and the problem formulation are described in Section.~\ref{sec:system}. The beam prediction including parameter estimation and hybrid beamforming, is described in Section.~\ref{sec:estimation}. The numerical results are given in Section.~\ref{sec:simulation}, and the conclusions are drawn in Section.~\ref{sec:conclusion}.

Notations: We use lowercase (uppercase) boldface $ \boldsymbol{A}(\boldsymbol{a}) $ to denote the vector (matrix), and $ a $ is a scalar. Calligraphy letter $ \mathcal{A} $ represents the set or the probability distribution. Superscripts $ (\cdot) $, $ (\cdot)^* $ and $ (\cdot)^H $ represent the transpose, the complex conjugate and the Hermitian transpose, respectively. $ \mathbb{E}\{\cdot\} $ denotes the expectation operator. $ \boldsymbol{I}_N $ denotes the an $ N\times N $ identity matrix, and $ \boldsymbol{n} \sim \mathcal{CN}(\boldsymbol{0}, \boldsymbol{I}_N) $ means $ \boldsymbol{n} $ is complex circularly-symmetric Gaussian distributed with zero mean and covariance $ \boldsymbol{I}_N $. $ |\cdot| $ is an absolute operator, $ \|\cdot\|_p $ denotes the $ \ell_p $ norm. $ \mathbb{R} $ and $ \mathbb{C} $ represent the real field and complex field, respectively.

\section{System Model and Problem Formulation}\label{sec:system}

\subsection{System Model}\label{secsub:system}

We consider a link-level multi-user (MU)-MIMO mmWave communication system composed of a BS and several MTs. The BS is equipped with $ N_{\textup{t}} $ antennas and $ N_{\textup{rf}} $ radio frequency (RF) chains, and the RF chains are fully connected with the antennas. The BS simultaneously serves $ N_{\textup{rf}} $ MTs, and each MT is equipped with $ N_{\textup{r}} $ antennas and $ 1 $ RF chain. In practice, both the analog transmitter precoder $ \boldsymbol{A}_{\textup{t}} $ and receiver precoder $ \boldsymbol{A}_{\textup{r}} $ are realized by the discrete Fourier transform (DFT) codebooks, i.e., $ \boldsymbol{A}_{\textup{t}, i} \in \big\{\mathcal{F}_{\textup{t}, j}|\forall j \in \{1, \cdots, N_{\textup{t}}\} \big\},\,\forall i \in \{1, \cdots, N_{\textup{{rf}}}\} $ and $ \boldsymbol{A}_{\textup{r}} \in \big\{\mathcal{F}_{\textup{r}, j}|\forall j \in \{1, \cdots, N_{\textup{r}}\} \big\} $. The received signal of MT $ u $ in the antenna field is represented as follows
\begin{equation}
\boldsymbol{y}_u = \boldsymbol{H}_u \boldsymbol{A}_{\textup{t}} \boldsymbol{D} \boldsymbol{s} + \boldsymbol{n}_u,
\end{equation}
where $ \boldsymbol{H} $ denotes the the channel matrix, $ \boldsymbol{D} \in \mathbb{C}^{N_{\textup{rf}} \times N_{\textup{rf}}} $ is and transmitter digital precoder, $ \boldsymbol{s} $ is the baseband signal, $ \boldsymbol{n} \sim \mathcal{CN}(\boldsymbol{0}, \sigma_n^2 \boldsymbol{I}_{N_{\textup{r}}}) $ is the additional noise, respectively. According to the mmWave channel model\footnote{We assume the BS and the MTs are (approximately) on the same horizontal plane, and thus the uniform linear array (ULA) is considered.}, $ \boldsymbol{H} $ is a sum of the contributions of $ K $ dominant paths, thus the discrete-time narrow-band channel matrix can be described as
\begin{equation}\label{equ:H}
\boldsymbol{H} = \sqrt{\frac{N_{\textup{t}} N_{\textup{r}}}{K}} \sum_{k=1}^{K} \alpha_k \boldsymbol{a}_{\textup{r}}(\phi_{k}^{\textup{r}}) \boldsymbol{a}_{\textup{t}}^H(\phi_{k}^{\textup{t}}),
\end{equation}
where $ \alpha_k $ is a complex channel gain of path $ k $, $ \phi^{\textup{r}} $ and $ \phi^{\textup{t}} $ are the azimuth angles of arrival (AoA) and departure (AoD), respectively. The array spacing is half of the carrier wavelength, and the array responses at the receiver and transmitter are respectively given as follows
\begin{equation}
\begin{split}
\boldsymbol{a}_{\textup{r}}(\phi) & = \frac{1}{\sqrt{N_{\textup{r}}}}[1, e^{j\pi\sin(\phi)}, \cdots, e^{j\pi (N_{\textup{r}}-1) \sin(\phi)}]^T,\\
\boldsymbol{a}_{\textup{t}}(\phi) & = \frac{1}{\sqrt{N_{\textup{t}}}}[1, e^{j\pi\sin(\phi)}, \cdots, e^{j\pi (N_{\textup{t}}-1) \sin(\phi)}]^T.\\
\end{split}
\end{equation}

Abundant prior knowledge is available in the scene of HSR, and they are beneficial to simplify the beam prediction. We summarize the prior knowledge point by point:
\begin{enumerate}
	\item The channel always contains a LoS path.\label{ass:1}
	\item The power of LoS path is much higher than the non-LoS (NLoS) paths, i.e., $ |\alpha_1|^2 \gg |\alpha_k|^2,\, k \neq 1 $.\label{ass:2}
	\item The MT moves along the track at some initial speed $ v $ and acceleration $ a $.\label{ass:3}
	\item The mapping between the AoD of the LoS path $ \phi $ and the corresponding spatial location projected on the x-axis $ x $ is a bijection, i.e., $ x \stackrel{\Phi}{\rightleftharpoons} \phi $.\label{ass:4}
\end{enumerate}
According to the prior knowledge \ref{ass:1}) and \ref{ass:2}), the channel function in \eqref{equ:H} can be further simplified as
\begin{equation}\label{equ:H_equ}
\begin{split}
\boldsymbol{H} & \approx \sqrt{N_{\textup{t}} N_{\textup{r}}} \alpha \boldsymbol{a}_{\textup{r}}(\phi^{\textup{r}}) \boldsymbol{a}_{\textup{t}}^H(\phi^{\textup{t}})\\
& = \sqrt{N_{\textup{t}} N_{\textup{r}}} \alpha \boldsymbol{a}_{\textup{r}}(\pi - \phi) \boldsymbol{a}_{\textup{t}}^H(\phi).\\
\end{split}
\end{equation}
Therefore, the channel matrix can be described by a parameter set of only two elements, i.e., $ \{\alpha, \phi\} $. 

\subsection{Problem Formulation}

\begin{figure}
	\centering
	\includegraphics[width=3.5in]{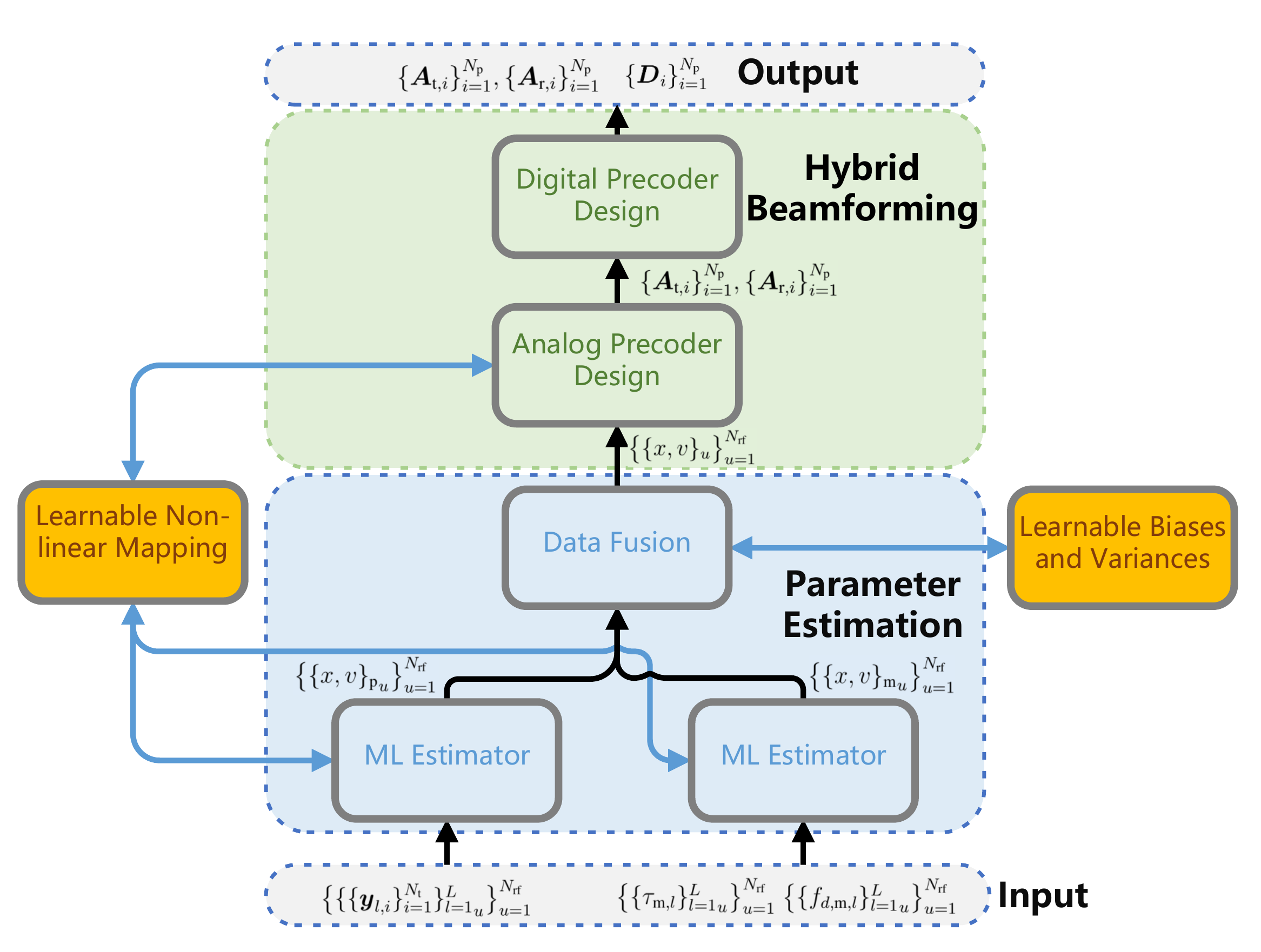}
	\caption{An illustration of beam prediction procedure. The parameter estimations and the analog precoder design are parallel carried out with respect to $ N_{\textup{rf}} $ MTs. The learning modules are labeled with orange color.}
	\label{fig:beam_prediction}
\end{figure}

As shown in Fig.~\ref{fig:beam_prediction}, in the first step of beam prediction procedure, the parameters of the parametric motion model such as projected location and speed, are required to be estimated by different observations. Specifically, the observations are carried out $ L $ times with fixed time interval $ \Delta t $. At each instant, the BS transmits all horizontal pilot beams. The first observation is the downlink pilot signal. Given the projected location $ x_l $ at instant $ l $, the received pilot signal $ \boldsymbol{y}_{l, i} $ of beam $ i $ can be written as
\begin{equation}
\begin{split}
\boldsymbol{y}_{l, i} & = \boldsymbol{H}_l \boldsymbol{A}_{\textup{t}, i} \boldsymbol{D} \boldsymbol{s}_{\textup{p}} + \boldsymbol{n}_{l, i}\\
& \approx \sqrt{N_{\textup{t}} N_{\textup{r}}} \alpha_l \underbrace{\Big(\boldsymbol{a}_{\textup{r}} \big(\Phi(x_l) - \pi \big) \boldsymbol{a}_{\textup{t}}^H \big(\Phi(x_l)\big) \Big) \boldsymbol{A}_{\textup{t}, i} \boldsymbol{D} \boldsymbol{s}_{\textup{p}}}_{\boldsymbol{z}_i(x_l)} + \boldsymbol{n}_{l, i},\\
\end{split}
\end{equation}
where $ \boldsymbol{s}_p $ is the pilot signal. Secondly, another observation is the measured communication delay $ \tau_{\textup{m}} $ and Doppler frequency $ f_{d, \textup{m}} $. At instant $ l $, the measurements are respectively given as follows
\begin{align}
\tau_{\textup{m}, l} & = \tau_{l} + n_{\tau, l},\\
f_{d, \textup{m}, l} & = f_{d, l} + n_{f_d, l},
\end{align}
where $ n_{\tau} \sim \mathcal{N}(0, \sigma_{\tau}^2) $ is the measurement error of communication delay with variance $ \sigma_{\tau}^2 $, and $ n_{f_d} \sim \mathcal{N}(0, \sigma_{f_d}^2) $ is the measurement error of Doppler frequency with variance $ \sigma_{f_d}^2 $. The variances are modeled by the range resolution and Doppler frequency resolution in radar theory. Besides, the residual carrier frequency is included in $ f_{d, \textup{m}} $. Therefore, variances are modeled as follows
\begin{align}
\sigma_{\tau}^2 & = \frac{c}{2 B},\\
\sigma_{f_d}^2 & = \frac{c}{2 f_c T_c} + k_{f_c} f_c,
\end{align}
where $ c $ is the light speed, $ B $ is the bandwidth, $ f_c $ is the carrier frequency, $ T_c $ is the integral time, and $ k_{f_c} $ is the residual carrier frequency ratio.

The acceleration is considered in the assumption \ref{ass:3}) in \ref{secsub:system}. To ensure the passengers' comfort, the absolute value of HSR acceleration is relatively small. Besides, we prove that accurate estimation of acceleration with limited observation interval and times is infeasible in Appendix~\ref{sec:appendix_a}. Therefore, the effects of HSR acceleration can be neglected, and we only estimate the projected location and speed of MTs. Particularly, the parameter estimation problem is: given a group of received downlink pilot signal $ \{\{\boldsymbol{y}_{l, i}\}_{i=1}^{N_{\textup{t}}}\}_{l=1}^L $ and measurements of communication delays $ \{\tau_{\textup{m}, l}\}_{l=1}^L $ and Doppler frequencies $ \{f_{d, \textup{m}, l}\}_{l=1}^L $, how to estimate the parameter set $ \{x, v\} $ of the MT, where $ x $, and $ v $ are respectively the projected location and speed at the final instant $ L $. 

\section{Beam Prediction}\label{sec:prediction}

\subsection{Parameter Estimation}\label{sec:estimation}

\subsubsection{Linear Tracks}

\begin{figure}
\centering
\subfigure[Linear track.]{
\includegraphics[width=3.0in]{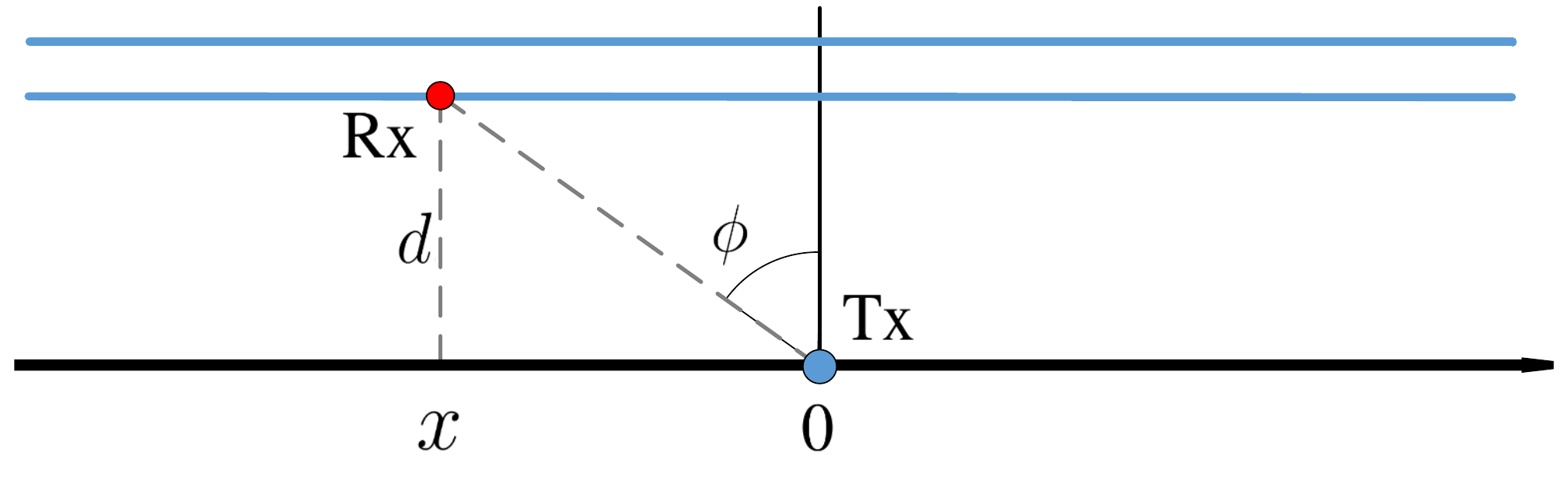}
\label{fig:lin_track}
}
\\
\subfigure[Non-linear track.]{		
\includegraphics[width=3.0in]{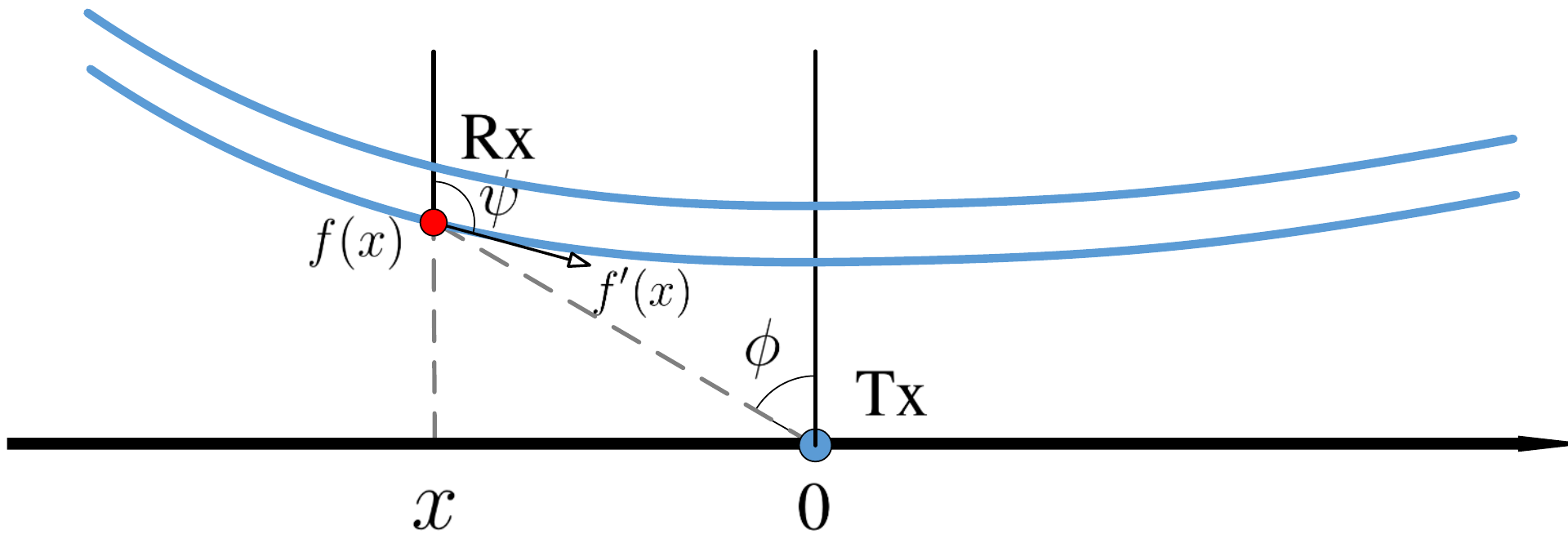}
\label{fig:nlin_track}
}
\caption{Illustrations of HSR tracks.}
\end{figure}

Firstly, we consider a simple case where the track is modeled as a straight line parallel to the x-axis with fixed distance $ d $, as shown in Fig.~\ref{fig:lin_track}. With the assumption of simplified uniform motion, the projected locations can be expressed as
\begin{equation}\label{equ:x_l}
x_l = x + (l-L) v \Delta t, \, \forall l.
\end{equation}
When the MT moves from left to right, speed $ v $ is regarded as positive, and vice versa. Therefore, the expression of function $ \Phi $ can be expressed as follows
\begin{equation}\label{equ:phi_linear}
\phi_l = \arctan(\frac{x_l}{d}).
\end{equation}
With parameter set $ \Theta = \big\{x, v, \{\alpha_l\}_{l=1}^{L}\big\} $, the posterior probability of received pilot signal is expressed as follows
\begin{equation}
p(\boldsymbol{y}_{l, i}; \Theta) = \frac{1}{\sqrt{2 \pi} \sigma_n}\exp\Big(-\frac{\big(\boldsymbol{y}_{l} - \alpha_l \boldsymbol{z}_i(x_l)\big)^H \big(\boldsymbol{y}_{l} - \alpha_l \boldsymbol{z}_i(x_l)\big)}{2 \sigma_n^2}\Big).
\end{equation}
Therefore, the overall posterior probability of a group of received pilot signals can be described as follows 
\begin{equation}
p(\{\{\boldsymbol{y}_{l, i}\}_{i=1}^{N_{\textup{t}}}\}_{l=1}^L; \Theta) = \prod_{l=1}^{L} \prod_{i=1}^{N_{\textup{t}}} p(\boldsymbol{y}_{l, i}; \Theta).
\end{equation}
To estimate the binary set $ \{x, v\} $ with a unique solution, the number of measurements must be larger than the number of elements, i.e., $ L \geq 2 $.

The prior of $ \Theta $ can be difficult to obtain, and the parameter set $ \Theta $ can be estimated by maximum likelihood (ML) criterion without prior of $ \Theta $. The posterior probability is non-convex with respect to multi-dimensional set $ \Theta $. Due to exponential computational complexity, exhaustive search in the high-dimensional parameter space is difficult and inefficient. To reduce the computational complexity, we propose to use coordinate descent method which updates the elements in the parameter set alternately and iteratively~\cite{2012Sparsity}. In the first iteration, the parameter elements are unknown, and the initial elements are obtained as follows
\begin{align}
x_l^{0} & = \arg \max_{x} p(\{\boldsymbol{y}_{l, i}\}_{i=1}^{N_{\textup{t}}}; x, \alpha_l = 1), \, \forall l,\label{equ:1_x0}\\
\alpha_l^{0} & = \frac{\sum_{i=1}^{N_{\textup{t}}} \boldsymbol{y}_{l, i}^H \boldsymbol{z}_{i}(x_l^{0})}{\sum_{i=1}^{N_{\textup{t}}} \boldsymbol{z}_{i}^H(x_l^{0}) \boldsymbol{z}_{i}(x_l^{0})}, \, \forall l,\label{equ:1_alpha0}\\
v^{0} & = \arg \max_{v} p(\{\{\boldsymbol{y}_{l, i}\}_{i=1}^{N_{\textup{t}}}\}_{l=1}^L; x^{0}, v, \{\alpha_l^{0}\}_{l=1}^{L}).\label{equ:1_v0}
\end{align}
The closed-form of $ \alpha $ is given in \eqref{equ:1_alpha0}. Due to the non-convexity of \eqref{equ:1_x0} and \eqref{equ:1_v0}, $ x $ and $ v $ are both derived by one-dimensional search. Similarly, the $ k $-th derivation is given as follows
\begin{align}
\alpha_l^{k} & = \frac{\sum_{i=1}^{N_{\textup{t}}} \boldsymbol{y}_{l, i}^H \boldsymbol{z}_{i}(x_l^{k-1})}{\sum_{i=1}^{N_{\textup{t}}}\boldsymbol{z}_{i}^H(x_l^{k-1}) \boldsymbol{z}_{i}(x_l^{k-1})}, \, \forall l,\label{equ:1_alphak}\\
x^{k} & = \arg \max_{x} p(\{\{\boldsymbol{y}_{l, i}\}_{i=1}^{N_{\textup{t}}}\}_{l=1}^L; x, v^{k-1}, \{\alpha_l^{k-1}\}_{l=1}^{L}),\label{equ:1_xk}\\
v^{k} & = \arg \max_{v} p(\{\{\boldsymbol{y}_{l, i}\}_{i=1}^{N_{\textup{t}}}\}_{l=1}^L; x^{k-1}, v, \{\alpha_l^{k-1}\}_{l=1}^{L}).\label{equ:1_vk}
\end{align}
Consider a linear track, the complete parameter estimation algorithm with the received pilot signals is given in \textbf{Algorithm}~\ref{alg:alg_1}.

\begin{algorithm}
	\caption{Linear track: parameter estimation with the received pilot signals (single MT).}
	\begin{algorithmic}[1]
		\STATE \emph{Input:} Received pilot signals $ \{\{\boldsymbol{y}_{l, i}\}_{i=1}^{N_{\textup{t}}}\}_{l=1}^L $, thresholds $ \{\theta_{\textup{th}, x}, \theta_{\textup{th}, v}\} $, maximum iteration $ K_{\textup{max}} $.
		\STATE \emph{Initialization:} Obtain $ \{x^0, \{\alpha^0_l\}_{l=1}^{L}, v^0\} $ by \eqref{equ:1_x0},  \eqref{equ:1_alpha0} and \eqref{equ:1_v0}, respectively.
		\FOR{$ k = 1 $ to $ K_{\textup{max}} $}
		\STATE $ x \gets x^{k-1} $, $ \{\alpha_l\}_{l=1}^{L} \gets \{\alpha^{k-1}_l\}_{l=1}^{L} $ and $ v \gets v^{k-1} $.
		\STATE Obtain $ \{x_l\}_{l=1}^{L} $ by \eqref{equ:x_l}.
		\STATE Obtain $ \{\alpha_l^{k}\}_{l=1}^{L} $, $ x^k $ and $ v^k $ by \eqref{equ:1_alphak}, \eqref{equ:1_xk} and \eqref{equ:1_vk}, respectively.
		\IF{$ |x - x^k| < \theta_{\textup{th}, x} $ and $ |v - v^k| < \theta_{\textup{th}, v} $}
		\STATE \textbf{break}
		\ENDIF
		\ENDFOR
		\STATE \emph{Output:} Estimated parameter set $ \{x, v\}_{\textup{p}} $.
	\end{algorithmic}
	\label{alg:alg_1}
\end{algorithm}

The parameter estimation by received pilot signals and measurements are carried out independently. According to the geometric relationship between the BS and MT in Fig.~\ref{fig:lin_track}, the derivation of $ \{\tau_l, {f_{d, l}}\} $ with respect to $ x_l $ can be described as follows
\begin{align}
\tau_l & = \frac{\sqrt{x_l^2 + d^2}}{c},\label{equ:tau_infer}\\
{f_{d, l}} & = \frac{2 f_c v'_l}{c}\nonumber\\
& = - \frac{2 f_c |v| \sin(\phi_l)}{c}\nonumber\\
& = - \frac{2 f_c |v| \sin(\arctan \frac{x_l}{d})}{c},\label{equ:fd_infer}
\end{align}
where $ v' $ is the speed component along the direction of LoS. With \eqref{equ:tau_infer} and \eqref{equ:fd_infer}, the posterior probabilities of the $ l $-th measurement can be respectively described as follows
\begin{align}
p(\tau_{\textup{m}, l}; \Theta) & = \frac{1}{\sqrt{2 \pi} \sigma_{\tau}} \exp \Big(- \frac{\big(\tau_{\textup{m}, l} - \tau_l\big)^2}{2 \sigma_{\tau}^2} \Big),\\
p(f_{d, \textup{m}, l}; \Theta) & = \frac{1}{\sqrt{2 \pi} \sigma_{f_d}} \exp \Big(- \frac{\big(f_{d, \textup{m}, l} - f_{d, l})^2}{2 \sigma_{f_d}^2} \Big).
\end{align}
The overall posterior probability of a group of measurements can be described as follows
\begin{equation}
\begin{split}
p(\{\tau_{\textup{m}, l}\}_{l=1}^{L}, \{{f_{d, \textup{m}, l}}\}_{l=1}^{L}; \Theta) & = \prod_{l=1}^{L} p(\tau_{\textup{m}, l}, f_{d, \textup{m}, l}; \Theta)\\
& = \prod_{l=1}^{L} p(\tau_{\textup{m}, l}; \Theta) p(f_{d, \textup{m}, l}; \Theta).
\end{split}
\end{equation}
Similarly, the estimation problem with given measurements can be solved by alternating iteration optimization. The initialization of the parameters $ x^0 $ and $ v^0 $ are respectively given as follows
\begin{align}
x^0 & = - \textup{sign}({f_{d, \textup{m}, L}})\sqrt{c^2\tau_{\textup{m}, L}^2 - d^2},\label{equ:2_x0}\\
v^0 & = - \frac{c {f_{d, \textup{m}, L}}}{2 f_c \sin(\arctan \frac{x^0}{d})},\label{equ:2_v0}
\end{align}
where $ \textup{sign}(\cdot) $ is the symbolic function. In iteration $ k $, the parameters are derived by one-dimensional search, and they are respectively given as follows
\begin{align}
x^{k} & = \arg \max_{x} p(\{\tau_{\textup{m}, l}\}_{l=1}^{L}, \{{f_{d, \textup{m}, l}}\}_{l=1}^{L}; x, v^{k-1}),\label{equ:2_xk}\\
v^{k} & = \arg \max_{v} p(\{\tau_{\textup{m}, l}\}_{l=1}^{L}, \{{f_{d, \textup{m}, l}}\}_{l=1}^{L}; x^{k-1}, v).\label{equ:2_vk}
\end{align}
Consider a linear track, the complete parameter estimation algorithm with the measurements is given in \textbf{Algorithm}~\ref{alg:alg_2}.

\begin{algorithm}
	\caption{Linear track: parameter estimation algorithm with the measurements (single MT).}
	\begin{algorithmic}[1]
		\STATE \emph{Input:} Observed $ \{\tau_{\textup{m}, l}\}_{l=1}^{L} $, $ \{f_{d, \textup{m}, l}\}_{l=1}^{L} $, thresholds $ \{\theta_{\textup{th}, x}, \theta_{\textup{th}, v}\} $, maximum iteration $ K_{\textup{max}} $.
		\STATE \emph{Initialization:} Obtain $ x^0 $ and $ v^0 $ by \eqref{equ:2_x0} and \eqref{equ:2_v0}, respectively.
		\FOR{$ k = 1 $ to $ K_{\textup{max}} $}
		\STATE $ x \gets x^{k-1} $ and $ v \gets v^{k-1} $.
		\STATE Obtain $ \{x_l\}_{l=1}^{L} $ by \eqref{equ:x_l}.
		\STATE Obtain $ x^k $ and $ v^k $ by \eqref{equ:2_xk} and \eqref{equ:2_vk}, respectively.
		\IF{$ |x - x^k| < \theta_{\textup{th}, x} $ and $ |v - v^k| < \theta_{\textup{th}, v} $}
		\STATE \textbf{break}
		\ENDIF
		\ENDFOR
		\STATE \emph{Output:} Estimated parameter set $ \{x, v\}_{\textup{m}} $.
	\end{algorithmic}
	\label{alg:alg_2}
\end{algorithm}

\subsubsection{Non-linear Tracks}

Secondly, we consider a more generalized case where the track is curved and the MTs move along the track with constant speed. The track is assumed to be a parallel straight line in the first case, but this assumption is not (strictly) true in many cases. To address this issue, we develop a data-driven and model-driven approach for parameter estimation. More concretely, the data-driven method is used to fit the non-linear track, and the model-driven method is used to estimate the parameter set by ML criterion.

As shown in Fig.~\ref{fig:nlin_track}, the track is modeled as an arbitrary projection distance function $ f(x) $ but follows the assumption \ref{ass:4}) in Section.~\ref{secsub:system}. The formula \eqref{equ:x_l} holds when the track is linear. When the track is non-linear, the solution of projected locations $ \{x_l\}_{l=1}^L $ derived by \eqref{equ:x_l} is replaced by
\begin{equation}\label{equ:x_l_non}
F(x_l, x) = (l-L) v \Delta t,
\end{equation}
where the function $ F(x_l, x) $ is defined as
\begin{equation}\label{equ:Fx}
F(x_l, x) = \int_{x_l}^{x} \sqrt{1+[f'(u)]^2} du,
\end{equation}
where $ f'(x) $ is the first-order derivative with respect to $ x $. When $ f(x) $ is known, we can only obtain an analytical solution of \eqref{equ:x_l_non} in most cases. Meanwhile, $ F(x_l, x) $ monotonically decreases with respect to $ x_l $. Therefore, the solution $ x_l $ of \eqref{equ:x_l_non} can be derived by binary searching in a lookup table. 

According to the principle of geometry, The initialization of the parameters $ x^0 $ and $ v^0 $ are respectively given as follows
\begin{align}
x^0 & = - \textup{sign}({f_{d, \textup{m}, L}})\sqrt{c^2\tau_{\textup{m}, L}^2 - f(x^0)^2},\label{equ:2_x0_non}\\
v^0 & = - \frac{c {f_{d, \textup{m}, L}}}{2 f_c \cos \big[\Phi(x^0) - \Psi(x^0, v^0)\big]},\label{equ:2_v0_non}
\end{align}
where the function $ \Phi $ and $ \Psi $ are respectively derived as follows
\begin{equation}\label{equ:phi_nonlinear}
\Phi(x) = \left\{
\begin{aligned}
& \arctan \frac{x}{f(x)}, & f(x) \leq 0,\\
& \arctan \frac{x}{f(x)} + \pi, & f(x) > 0,
\end{aligned}
\right.
\end{equation}
\begin{equation}\label{equ:psi_nonlinear}
\Psi(x, v) = \left\{
\begin{aligned}
& \arctan \frac{1}{f'(x)}, & v f'(x) \leq 0,\\
& \arctan \frac{1}{f'(x)} + \pi, & v f'(x) > 0.
\end{aligned}
\right.
\end{equation}
Both the formulas \eqref{equ:2_x0_non} and \eqref{equ:2_v0_non} are transcendental equations. Formula \eqref{equ:2_x0_non} can be solved with a numerical solution $ x^0 $. When $ x^0 $ is given, the $ \Psi $ in formula \eqref{equ:psi_nonlinear} is only related to the symbolic character of $ v $, thus \eqref{equ:2_v0_non} can be easily solved by a binary try. The expressions of $ \tau_l $ and $ f_{d, l} $ at projected location $ x_l $ can be respectively derived as follows
\begin{align}
\tau_l & = \frac{\sqrt{x_l^2+f^2(x_l)}}{c},\label{equ:tau_non}\\
f_{d, l} & = -\frac{2 f_c |v|}{c} \cos \big[\Phi(x_l) - \Psi(x_l, v)\big].\label{equ:f_d_non}
\end{align}
It is easy to prove that the Doppler frequency function \eqref{equ:fd_infer} is a special case of the generalized formula \eqref{equ:f_d_non}. The parameter estimation algorithm with the received pilot signals considering a non-linear track is similar to \textbf{Algorithm}~\ref{alg:alg_1} which considers a linear track. We highlight the differences between the linear and non-linear cases as follows
\begin{itemize}
	\item Consider a linear track, the location set $ \{x_l\}_{l=1}^L $ is computed by \eqref{equ:x_l}. Consider a non-linear track, the location set $ \{x_l\}_{l=1}^L $ is instead derived by \eqref{equ:x_l_non} in the binary search method.
	\item Consider a linear track, the initialization of $ \{x, v\} $ is derived by \eqref{equ:2_x0} and \eqref{equ:2_v0}. Consider a non-linear track, the initialization of $ \{x, v\} $ is instead derived by \eqref{equ:2_x0_non} and \eqref{equ:2_v0_non}.
	\item Consider a linear track, the set $ \{\tau_l, f_{d, l}\}_{l=1}^L $ is derived by \eqref{equ:tau_infer} and \eqref{equ:fd_infer}. Consider a non-linear track, the set $ \{\tau_l, f_{d, l}\}_{l=1}^L $ is instead derived by \eqref{equ:tau_non} and \eqref{equ:f_d_non}.
\end{itemize}
In addition to these differences, both the parameter estimation algorithms (with the received pilot signals and with the measurements) in the non-linear case are the same as these in the linear case.

\subsubsection{Function Fitting}\label{sec:fun_fit}

The derivations in the non-linear case are obtained under the condition that the track function $ f(x) $ is known. In practice, the deterministic function $ f(x) $ is unknown. Hence, $ f(x) $ can be learned by some parametric function $ g(x; \Theta_g) $ in a data-driven manner, where $ \Theta_g $ is the network parameter set. Labeled data set can be offline collected by geometric measurements, such as aerial photography and satellite photography.

According to the universal approximation theorem, using sufficient hidden computing units, the multi-layer perceptron (MLP) can approximate the track with arbitrary accuracy. However, a regular MLP usually has redundant parameters and lacks interpretability. We propose to construct a contribution of $ N_x $ piece-wise linear functions to fit the function $ f(x) $, and the parametric function $ g(x; \Theta_g) $ is defined as follows
\begin{equation}\label{equ:gx}
g(x; \Theta_g) = \sum_{i=1}^{N_x} \Big[\frac{y_{i+1}-y_{i}}{x_{i+1}-x_{i}}(x-x_{i}) + y_{i}\Big] \Pi(x, x_{i}, x_{i+1}),
\end{equation}
where $ \Theta_g = \{y_{i}\}_{i=1}^{N_x+1} $ denotes the learnable parameter set, and the rectangular window function $ \Pi(x, a, b) $ is defined as follows
\begin{equation}\label{equ:Pi}
\Pi(x, a, b) =
\begin{cases}
1, & x \geq a \,\, \textup{and} \,\, x \leq b,\\
0, & \textup{otherwise},\\
\end{cases}
\end{equation}
and the range $ x \in [-r_{\textup{max}}, r_{\textup{max}}] $ is equally divided into $ N_x $ pieces where $ r_{\textup{max}} $ is the BS maximum service radius. The expression of $ x_i $ is given as follows
\begin{equation}\label{equ:x_i}
x_i = - r_{\textup{max}} + (i-1)\Delta x,\,\forall i \in \{1,\cdots, N_x+1\},
\end{equation}
where $ \Delta x = \frac{2 r_{\textup{max}}}{N_x} $ denotes the range interval. Apparently, function $ g(x) $ is continuous on the support set of $ x $. The training is carried out in a SL-enabled approach. The cost function\footnote{Consider the convenience of theoretical analysis, we use the mean absolute error (MAE) as the measure. In practice, the mean square error (MSE) is also feasible.} is defined as
\begin{equation}\label{equ:loss_fitting}
L(\Theta_g) = \mathbb{E}_{x \sim \mathcal{X}}\Big\{\big\|f_n(x) - g(x; \Theta_g) \big\| \Big\},
\end{equation}
where $ f_n(x) $ means $ f(x) $ is observed with additional Gaussian noise. The parameter set $ \Theta_g $ is iteratively updated by mini-batch gradient descent (MBGD) until convergence. Due to the determinacy of the tracks, the deployed learned function $ g(x; \Theta_g) $ does not require any online fine-tuning or periodic update. The corresponding analysis about the expected loss in \eqref{equ:loss_fitting} is presented in Appendix~\ref{sec:appendix_b}.

\subsubsection{Data Fusion}

Ignoring the effects of acceleration, the parameter set $ \{x, v\} $ is the sufficient statistics for the following beam prediction. In Section.~\ref{sec:estimation} we already have two estimated parameter sets which are derived from different observations independently. According to the statistical theory, there exists an optimal estimation from a group of independent observations. Meanwhile, the optimality is only guaranteed when the following assumptions hold true:
\begin{itemize}
	\item The estimated variables follow Gaussian distribution;
	\item The estimations are unbiased;
	\item The variances of estimated variables are known.
\end{itemize}
Firstly, due to the complexity and randomness of the practical wireless communication scenario, such as imperfect hardware and inaccurate models, the above assumptions cannot hold true and thus the performance gap between the theoretical and practical results is conspicuous. Secondly, there exists a potential mapping function between the projected location and the estimation accuracy. For example, when the MTs are far away from the BS, the estimation variance by the received pilot signals is very large due to high path loss and limited angular resolution, and vice versa. This indicates that using this mapping function can improve the estimation precision.

\begin{figure*}
	\centering
	\includegraphics[width=6.0in]{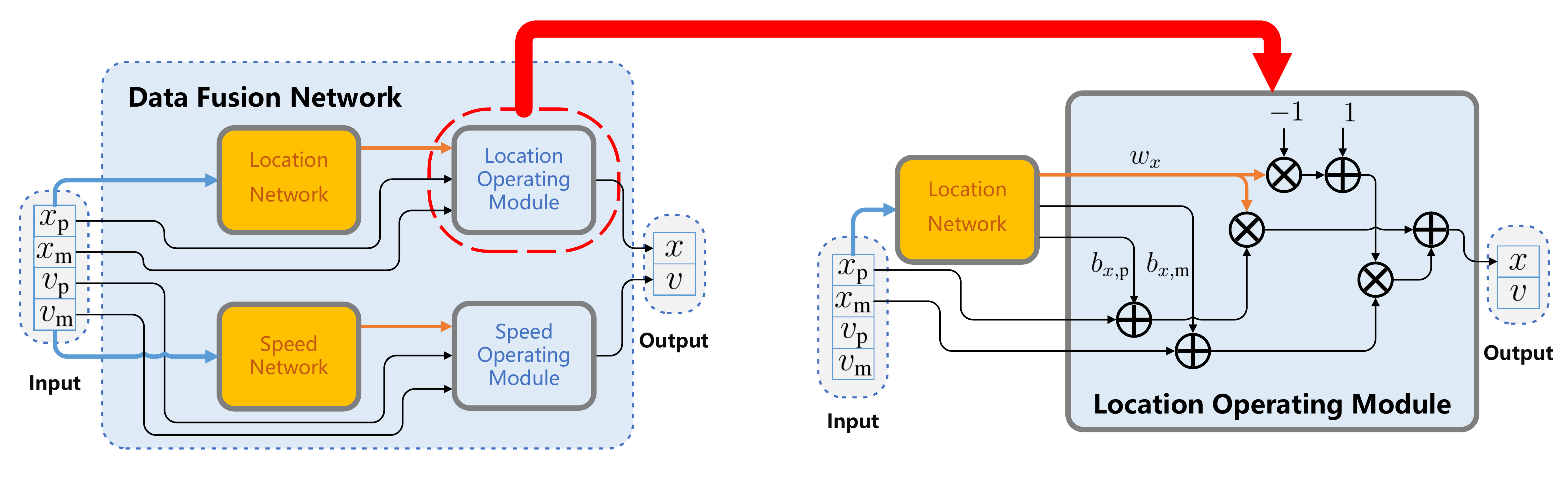}
	\caption{An illustration of data fusion network. The learning modules are labeled with orange color.}
	\label{fig:data_fusion}
\end{figure*}

Out of these two motivations, we propose to develop a data-driven data fusion method. To distinguish the estimation results, we mark the parameter set derived by the received pilot signals as $ \{x, v\}_{\textup{p}} $, and similarly mark the parameter set derived by the measurements as $ \{x, v\}_{\textup{m}} $. As shown in Fig.~\ref{fig:data_fusion}, $ \{x, v\}_{\textup{p}} $ and $ \{x, v\}_{\textup{m}} $ are then concatenated as the input of the NN model $ h(\cdot; \Theta_h) $, where $ \Theta_h = \{\Theta_x, \Theta_v\} $ denotes the network parameter set which is composed by the parameters of location network $ h_x $ and speed network $ h_v $. The topologies of the two networks are the same, and each network is composed by a weight sub-network and a bias sub-network as illustrated in Table~\ref{tab:network}. The notation 'BN' denotes batch normalization (BN), notation 'ReLU' denotes rectified linear unit (ReLU), and integer is computation unit number of this layer. The expressions of the two networks are respectively written as
\begin{align}
\big\{w_x, b_{x, \textup{m}}, b_{x, \textup{p}}\} & = h_x\Big(\big\{\{x, v\}_{\textup{p}}, \{x, v\}_{\textup{m}}\big\}; \Theta_x\Big),\label{equ:hx_fusion}\\
\big\{w_v, b_{v, \textup{m}}, b_{v, \textup{p}}\} & = h_v\Big(\big\{\{x, v\}_{\textup{p}}, \{x, v\}_{\textup{m}}\big\}; \Theta_v\Big).\label{equ:hv_fusion}
\end{align}
In principle, the network $ h $ learns to estimate the variances and offsets of the input estimations implicitly, assigns the weights $ \mathcal{W} = \{w_x, w_v\} $ and biases $ \mathcal{B} = \{b_{x, \textup{p}}, b_{x, \textup{m}}, b_{v, \textup{p}}, b_{v, \textup{m}}\} $ for the input estimations. The proposed data fusion network shares the same principle as that of the well-studied attention networks~\cite{attention}, which also adjust the weights by the input features. The output of the network is $ \{x, v\} $. Inspired by the model-based estimation method, the output estimations are respectively derived as
\begin{align}
x|_{\Theta_x} & = w_x (x_{\textup{p}} - b_{x, \textup{p}}) + (1 - w_x) (x_{\textup{m}} - b_{x, \textup{m}}),\label{equ:x_fusion}\\
v|_{\Theta_v} & = w_v  (v_{\textup{p}} - b_{v, \textup{p}}) + (1 - w_v) (v_{\textup{m}} - b_{v, \textup{m}}).\label{equ:v_fusion}
\end{align}
Compared to the regular NNs, the proposed data fusion network is light-weighted, inherently against over-fitting, and have a good interpretability.

The training can be realized in an open-loop manner, i.e., the network $ h $ is trained with prepared labeled data by SL. The training procedure is similar to the function fitting in Section.~\ref{sec:fun_fit}, and the cost function is defined as
\begin{equation}\label{equ:loss_fusion}
L(\Theta_h) = \mathbb{E}_{\{x, v\} \sim \mathcal{D}}\Big\{\big(x|_{\Theta_x} - x_{\textup{tar}} \big)^2 + \big(v|_{\Theta_v} - v_{\textup{tar}} \big)^2 \Big\},
\end{equation}
where the subscript $ (\cdot)_{\textup{tar}} $ denotes the labeled data. When some term dominates the overall cost function, theoretically the loss of other terms can rise. Meanwhile, we have observed that the loss of the other terms grows slowly, even with a small training set. Actually, the domination rarely occurs in practical problems. Therefore, \eqref{equ:loss_fusion} is formulated as a sum. The parameter set $ \Theta_h $ is iteratively updated by MBGD method until convergence. Besides, the close-loop training can be carried out by reinforcement learning.

\begin{table}
	\centering 
	\setlength{\tabcolsep}{0.5mm}{
	\scriptsize
	\centering 
	\caption{Topologies of the sub-networks.} 
	\begin{tabular}{ccc} 
		\toprule  
		& Weight sub-network & Bias sub-network \\
		\midrule
		Output layer & sigmoid, $ 1 $ & linear, $ 2 $\\
		BN layer & $ \backslash $ & $ \backslash $\\
		Hidden layer $ 2 $ & ReLU, $ 32 $ & ReLU, $ 32 $\\
		BN layer & $ \backslash $ & $ \backslash $\\
		Hidden layer $ 1 $ & ReLU, $ 32 $ & ReLU, $ 32 $\\
		Input layer & linear, $ 4 $ & linear, $ 4 $\\
		\bottomrule 
	\end{tabular}
	\label{tab:network}}
\end{table}

\subsection{Hybrid Beamforming}\label{sec:beam_mapping}

\begin{figure}
	\centering
	\includegraphics[width=3.0in]{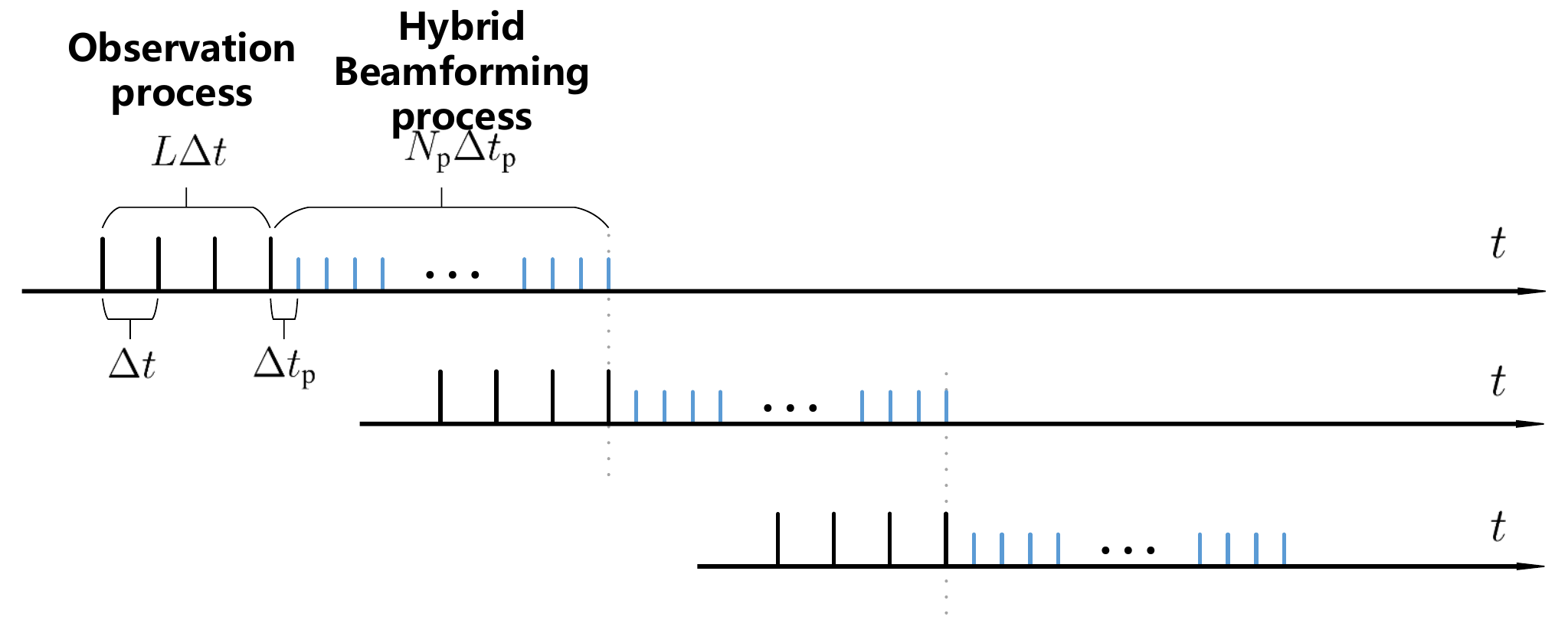}
	\caption{An illustration of the observation process and hybrid beamforming process along time.}
	\label{fig:problem_illustration}
\end{figure}

In our proposed beam prediction procedure, the high-dimensional beam prediction problem is equivalently transformed into a low-dimensional parameter estimation problem and a cascaded hybrid beamforming problem. In the hybrid beamforming, the hybrid precoders in a future time are predicted by the parameter set $ \{\{x, v\}_u\}_{u=1}^{N_{\textup{rf}}} $. As shown in Fig.~\ref{fig:problem_illustration}, the time granularity is $ \Delta t_{\textup{p}} $, and the number of predict instants is $ N_{\textup{p}} $. Therefore, the period of hybrid beamforming is $ N_{\textup{p}} \Delta t_{\textup{p}} $.

\subsubsection{Transmitter Analog Precoder and Receiver Combiner}\label{sec:analog}

Firstly, we consider a linear track and the predicted projected location of MT $ u $ at instant $ i $ is given as 
\begin{equation}\label{equ:x_mapping_lin}
x_{i, u} = x_u + v_u i \Delta t_{\textup{p}}.
\end{equation}
The corresponding AoD can be derived by \eqref{equ:phi_linear}. Secondly, consider a non-linear track, the derivations of projected location is rewritten as
\begin{equation}
F(x_u, x_{i, u}) = v_u i \Delta t_{\textup{p}},
\end{equation}
and the corresponding AoD is instead derived by \eqref{equ:phi_nonlinear}. The receiver combiner and the transmitter analog precoder are respectively derived by
\begin{align}
{\boldsymbol{a}_{\textup{r}, i, u}} & = \arg \min_{\forall j \in \{1, \cdots, N_{\textup{r}}\}} \|{\mathcal{F}_{\textup{r}, j}} - \boldsymbol{a}_{\textup{r}}(\phi_{i, u})\|^2,\label{equ:ar}\\
{\boldsymbol{a}_{\textup{t}, i, u}} & = \arg \min_{\forall j \in \{1, \cdots, N_{\textup{t}}\}} \|{\mathcal{F}_{\textup{t}, j}} - \boldsymbol{a}_{\textup{t}}(\phi_{i, u})\|^2.\label{equ:at}
\end{align}

\subsubsection{Digital Precoder}\label{sec:digital}

To simply the description, we take one instant of beam prediction for example. The digital precoding matrix is composed of $ N_{\textup{rf}} $ precoders, i.e., $ \boldsymbol{D}_i = [\boldsymbol{v}_{i, 1}, \cdots, \boldsymbol{v}_{i, N_{\textup{rf}}}] $. The equivalent low-dimensional channel is obtained as $ \boldsymbol{h}_{i, u} = \big(\boldsymbol{a}_{\textup{r}, i, u}^H \boldsymbol{H}_{i, u} \boldsymbol{A}_{\textup{t}, i}\big)^H $ which is obtained by CSI reference signal (CSI-RS) at the BS. We adopt a classical linear MMSE to derive the transmitter digital precoder as follows
\begin{equation}\label{equ:mmse}
\boldsymbol{D}_i = \xi \boldsymbol{H}_i\Big(\boldsymbol{H}_i^H \boldsymbol{H}_i + \sigma_n^2 \boldsymbol{I}_{N_{\textup{rf}}}\Big)^{-1}
\end{equation}
where $ \boldsymbol{H}_i = [\boldsymbol{h}_{i, 1}, \cdots, \boldsymbol{h}_{i, N_{\textup{rf}}}]^T $, $ \xi $ is a factor to control the BS maximum transmit power.

\subsubsection{Hybrid Precoding}\label{sec:hybrid}

The complete hybrid beamforming procedure with a linear track is given in \textbf{Algorithm}~\ref{alg:alg_3}. The procedure with a non-linear track is similar to \textbf{Algorithm}~\ref{alg:alg_3}. We highlight the differences between the linear and non-linear cases as follows
\begin{itemize}
	\item Consider a linear track, the projected location $ x_{i, u} $ is derived by \eqref{equ:x_mapping_lin}, and the corresponding AoD $ \phi_{i, u} $ is obtained by \eqref{equ:phi_linear}. Consider a non-linear track, $ x_{i, u} $ is derived by \eqref{equ:x_mapping_lin}, and $ \phi_{i, u} $ is obtained by \eqref{equ:phi_nonlinear}.
\end{itemize}
All the other steps are the same of those in the linear case, and then the hybrid beamforming procedure in the non-linear case is obtained. The hybrid precoders of different predicted instants can be parallel carried out.

\begin{algorithm}
\caption{Linear tracks: hybrid beamforming procedure (single instant).}
\begin{algorithmic}[1]
	\STATE \emph{Input:} Estimated parameter set $ \{\{x, v\}_u\}_{u=1}^{N_{\textup{rf}}} $.
	\STATE Obtain the projected location set $ \{x_{i, u}\}_{u=1}^{N_{\textup{rf}}} $ by \eqref{equ:x_mapping_lin}, and the corresponding AoD set $ \{\phi_{i, u}\}_{u=1}^{N_{\textup{rf}}} $ by \eqref{equ:phi_linear}.
	\STATE Respectively obtain the receiver combiner $ \boldsymbol{A}_{\textup{r}, i} = [\boldsymbol{a}_{\textup{r}, i, 1}, \cdots, \boldsymbol{a}_{\textup{r}, i, N_{\textup{rf}}}] $ and the transmitter analog precoder $ \boldsymbol{A}_{\textup{t}, i} = [\boldsymbol{a}_{\textup{t}, i, 1}, \cdots, \boldsymbol{a}_{\textup{t}, i, {N_{\textup{rf}}}}] $.
	\STATE Obtain the equivalent channel matrix $ \boldsymbol{H}_i = [\boldsymbol{h}_{i,1}, \cdots, \boldsymbol{h}_{i, {N_{\textup{rf}}}}]^T $, and derive the transmitter digital precoder $ \boldsymbol{D}_i $ by \eqref{equ:mmse}.
	\STATE \emph{Output:} The receiver combiner $ \boldsymbol{A}_{\textup{r}, i} $ and transmitter analog precoder $ \boldsymbol{A}_{\textup{t}, i} $, and the digital precoder $ \boldsymbol{D}_i $.
	\end{algorithmic}
	\label{alg:alg_3}
\end{algorithm}

\subsection{Implement}

Consider a non-linear track, the non-linear mapping module and the data fusion module are assumed to be off-line trained, and fine-tuned online. Finally, we summarize the implement procedure of beam prediction in Table~\ref{table:implement}.

\begin{table}[]
	\centering
	\setlength{\tabcolsep}{0.5mm}{
	\scriptsize
	\caption{Implement of beam prediction}
	\begin{tabular}{p{9cm}}
		\hline\hline
		\textbf{Initialization:}\\ The non-linear mapping module and the data fusion module.\\
		\hline
		\textbf{Observation process:}\\ $ 1) $ The BS transmits pilot signals to the MT.\\ 
		$ 2) $ The MT estimates the Doppler frequencies and ToAs, then feedbacks the received pilot signals, Doppler frequencies and ToAs to the BS.\\ 
		$ 3) $ The BS derives the final estimation result with feedbacks.\\
		\hline
		\textbf{Hybrid beamforming process:}\\ $ 1) $ The BS predicts the BS analog precoder and MT combiner with the final estimation result.
		\\ 
		$ 2) $ The BS transmits the MT combiner to the MT.\\ 
		$ 3) $ The BS transmits the data signals to the MTs with hybrid beamforming, and the MTs receive the signals with combiners.\\ 
		\hline\hline
	\end{tabular}
	\label{table:implement}}
\end{table}

\section{Simulation Results}\label{sec:simulation}

\subsection{System Configurations}

In this section, we present the simulation results to demonstrate the performance of the proposed learning-aided beam prediction scheme. Generally, the simulated mmWave channel in HSR is modeled as UMa LoS in $ 3 \textup{GPP} $ TR $ 38.901 $, and the wireless communication configurations are listed in Table~\ref{tab:sim_configuration}. The BS has $ 3 $ sectors and each sector covers $ 120^{\circ} $ range. Each MT has $ 3 $ panels (left, back, right). The speeds of MTs on board are modeled to follow Laplacian distribution, and the acceleration is also considered as an uncertain factor. Besides, the geometry of the established HSR scenario and the setting of the training beams/measurements~\cite{beam_enhance, 38901, Ericsson} are given in Table~\ref{tab:sim_configuration}. The BA/T is regarded as benchmark and only the horizontal beam alignment is considered. The BS/MT tracks 3 Tx/Rx beams (current, left and right) in each BA/T with period being $ 10 \; \textup{ms} $.

\begin{table*}
	\centering 
	\setlength{\tabcolsep}{0.5mm}{
	\scriptsize
	\caption{Simulation Configurations} 
	\begin{tabular}{cccccc} 
		\toprule
		Name & Value & Name & Value & Name & Value \\ 
		\midrule
		Scenario & UMa LoS & MT speed variance & $ 18 \; (\textup{km/h})^2 $ & Carrier Frequency $ f_c $ & $ 30 \; \textup{GHz} $\\
		MT acceleration variance & $ 0.1 \; (\textup{m/s}^2)^2 $ & Bandwidth $ B $ & $ 80 \; \textup{MHz} $ & HSR speed & $ 256 \; \textup{km/h} $\\
		Noise power spectral density & $ -174 \; \textup{dBm/Hz} $ & BS antenna number $ N_{\textup{t}} $ & $ 8 $ & BS maximum transmit power $ P_{\textup{t}, \textup{max}} $ & $ 30 \; \textup{dBm} $\\
		MT antenna number $ N_{\textup{r}} $ & $ 4 $ & Half inter site distance $ r_{\textup{max}} $ & $ 100 \; \textup{m} $ & prediction time duration & $ 1.25 \; \textup{s} $\\	
		Minimum BS to MT Distance $ d $ & $ 11 \; \textup{m} $ & Prediction time granularity $ \Delta t_{\textup{p}} $ & $ 1.25 \; \textup{ms} $ & Integral time $ T_c $ & $ 12.5 \; \textup{ms} $\\
		Observation period $ \Delta t $  & $ 100 \; \textup{ms} $ & Residual carrier frequency ratio $ k_{f_c} $ & $ 1 \; \textup{ppm} $ & Observation times $ L $ & $ 3 $\\		
		\bottomrule
	\end{tabular}
	\label{tab:sim_configuration}}
\end{table*}

\subsection{Data Fusion}

\subsubsection{Linear Tracks}

\begin{figure}
	\centering
	\subfigure[MSE of location estimation.]{
		\includegraphics[width=3.0in]{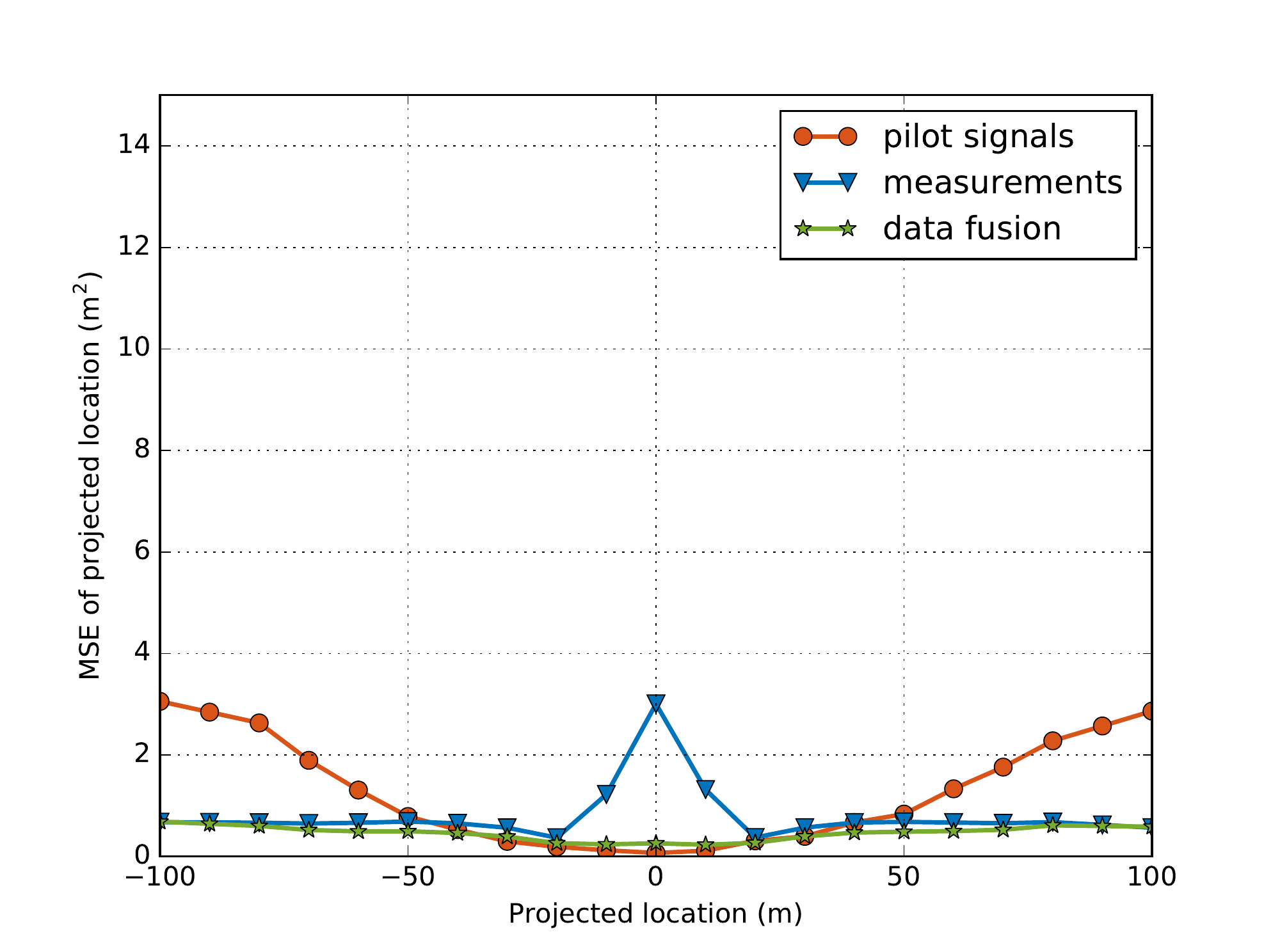}
		\label{fig:location_error}
	}
	\subfigure[MSE of velocity estimation.]{		
		\includegraphics[width=3.0in]{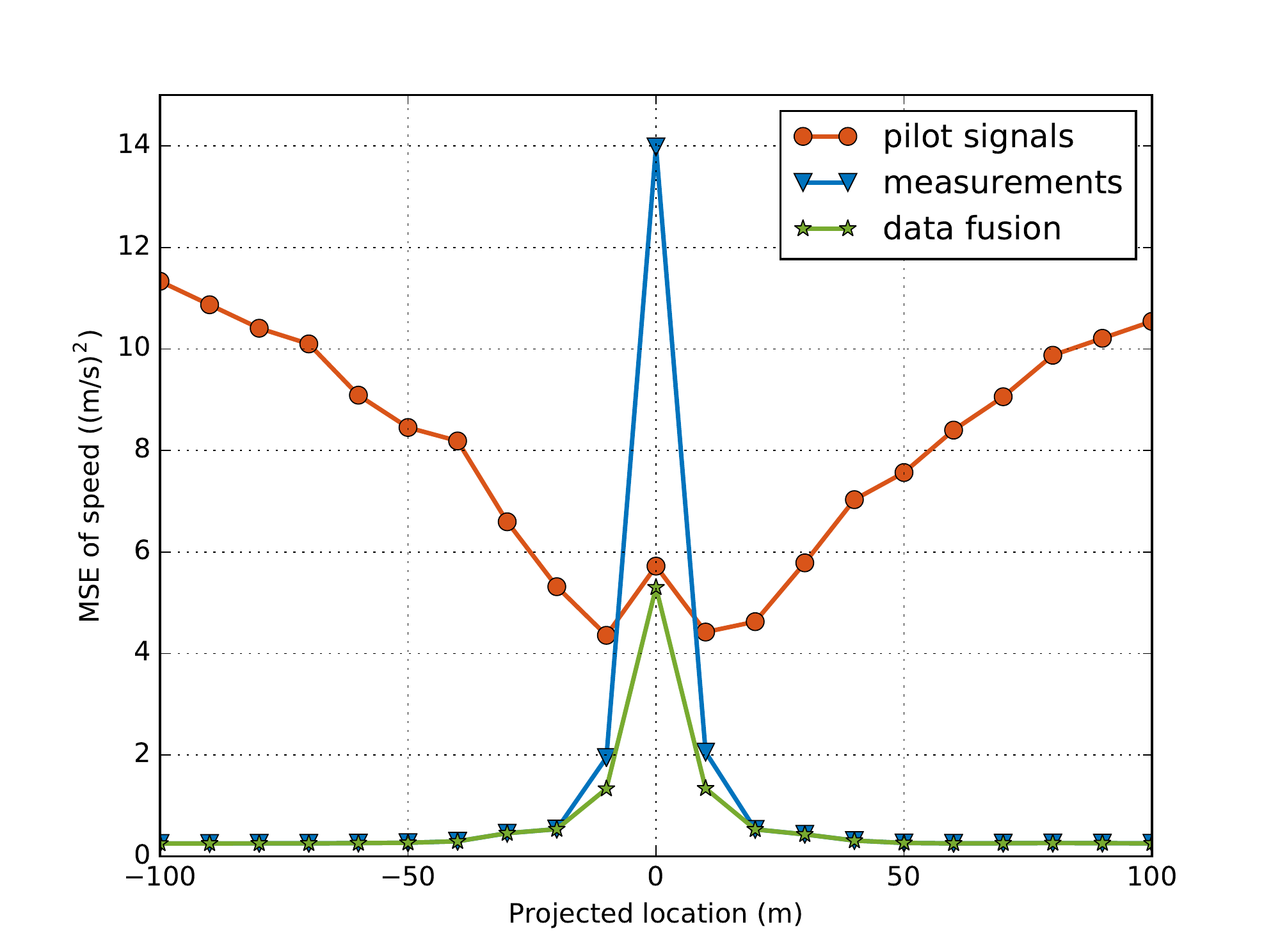}
		\label{fig:velocity_error}
	}
	\caption{Linear tracks: the MSE versus projected location $ x $.}
\end{figure}

As shown in Fig.~\ref{fig:location_error}, the location estimation accuracy by measurements is high when the MTs are far away from the BS, while the accuracy is low when the MTs are located adjacent to the BS. This phenomenon is caused by the lack of prior knowledge on the MT speed. The communication delay contains the information of distance between the BS and the MT, Whether the MT is located at right or left side of BS, however, cannot be inferred from the delay. Besides, we use symbolic character of Doppler frequency to discriminate the MT speed direction in \eqref{equ:2_x0}. However, the estimation performance cannot be improved especially when the measured Doppler frequency is significantly noised or the projected speed component is sharply reduced. Therefore, as shown in Figs.~\ref{fig:location_error} and \ref{fig:velocity_error}, neither the location and the speed estimations can be accurate in this range. 

Meanwhile, the estimations derived by the received pilot signals becomes more accurate when the BS to MT distance is reduced, because the corresponding path loss is reduced and the SNR of the signals is increased. Besides, the AoD of the MT is also easy to be distinguished in this range. When the MT moves far away form the BS, the variances of estimation sharply increased\footnote{The illustrated MSE curves are regularized by a maximum value being $ 30 $.}.

Generally, the estimation accuracies by measurements are more accurate than that by received pilot signals when the MTs are away from the BS, but the estimation accuracies by received pilot signals are more accurate than that by measurements when the MTs are around the BS. In a data-driven manner, our proposed data fusion method has the highest accuracy with respect to the projected location $ x $, both in the estimations of location and speed. The validity of the proposed method is verified by the simulation results, which also indicate that the function $ h $ have learned a weight function $ w(\cdot) $ with respect to $ x $.

\subsubsection{Non-linear Tracks}

\begin{figure}
	\centering
	\subfigure[MSE of location estimation.]{
		\includegraphics[width=3.0in]{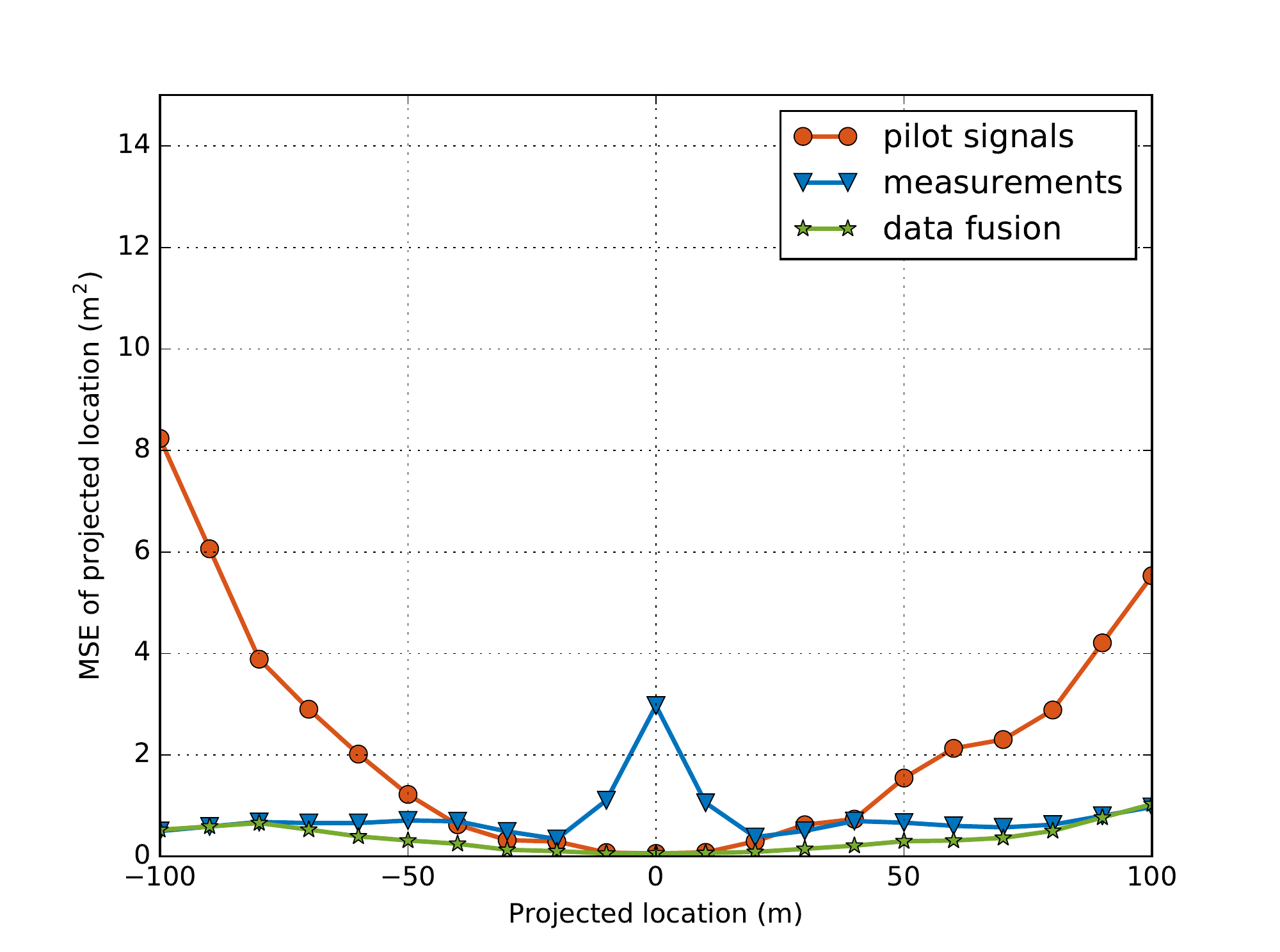}
		\label{fig:location_error_non}
	}
	\subfigure[MSE of velocity estimation.]{		
		\includegraphics[width=3.0in]{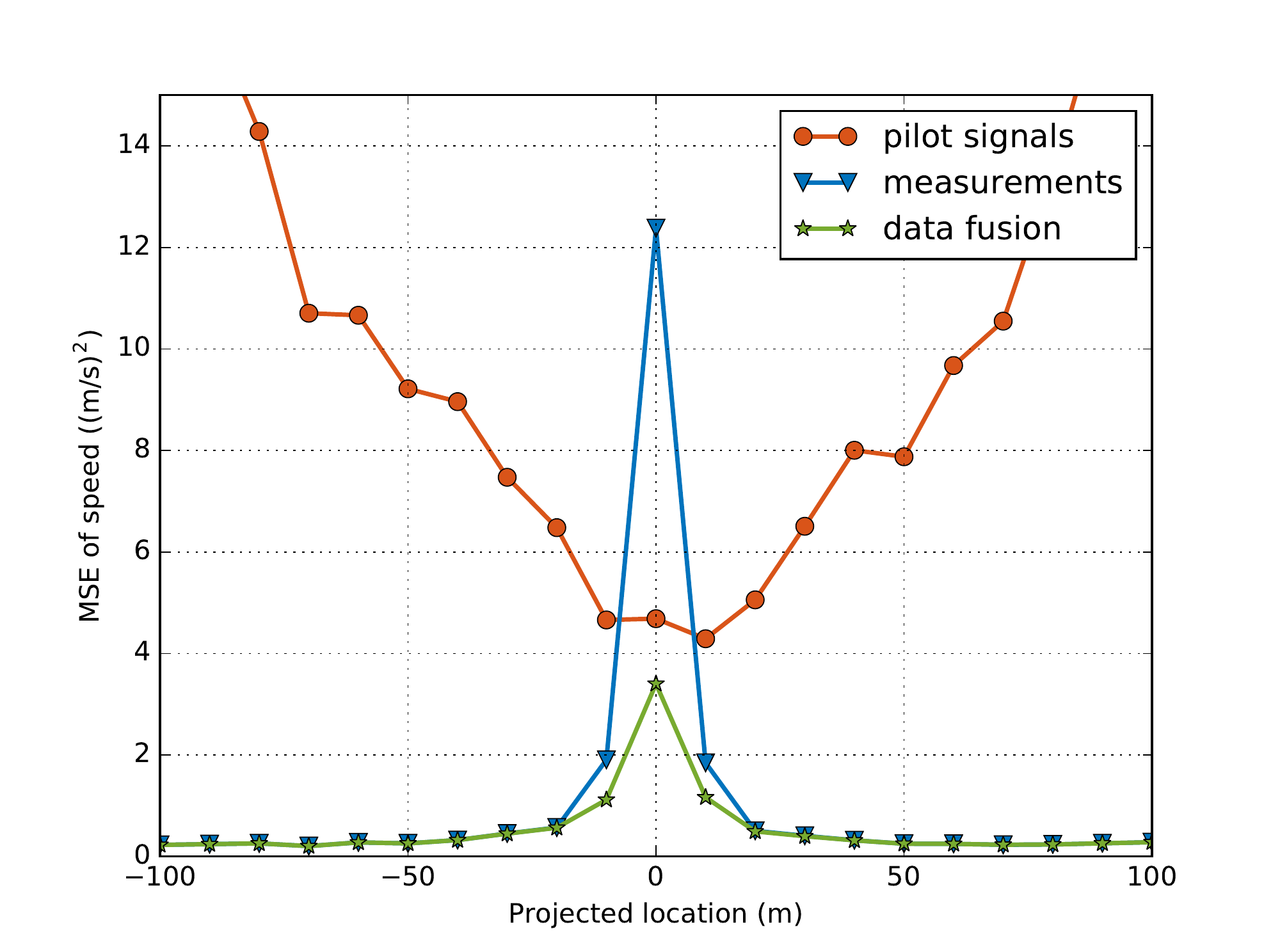}
		\label{fig:velocity_error_non}
	}
	\caption{Non-linear tracks (non-linear estimations): the MSE versus projected location $ x $.}
\end{figure}

In this part, we consider a more complex case where the track is modeled as a non-linear function, namely $ f(x) = \big(\frac{6}{200}\big)^2 (x-5)^2 + 11 $. Firstly, we demonstrate the estimation results with the non-linear track in Figs.~\ref{fig:location_error_non} and \ref{fig:velocity_error_non}. We compare Figs.~\ref{fig:location_error} and \ref{fig:location_error_non}, Figs.~\ref{fig:velocity_error} and \ref{fig:velocity_error_non}, and we have found that the trends of estimation variance with linear and non-linear tracks are similar. 

Using linear estimation algorithms described in \textbf{Algorithms}~\ref{alg:alg_1} and \ref{alg:alg_2}, the estimation results with the non-linear track are shown in Figs.~\ref{fig:location_error_non} and \ref{fig:velocity_error_non}. We compare the estimations with and without non-linear correction, and we have found that the estimation performance is significantly improved with non-linear correction, which verifies the effectiveness of our proposed non-linear mapping module. We have also noticed the data-driven method is significantly better than the primary two estimators in Figs.~\ref{fig:location_error_non} and \ref{fig:velocity_error_non}. This is mainly because the estimators become biased due to the model mismatch, and the data fusion module can reduce these biases and improve the estimation performance to some extent. Additionally, our proposed data fusion method also has the highest accuracy when the track is non-linear.

\begin{figure}
	\centering
	\subfigure[MSE of location estimation.]{
		\includegraphics[width=3.0in]{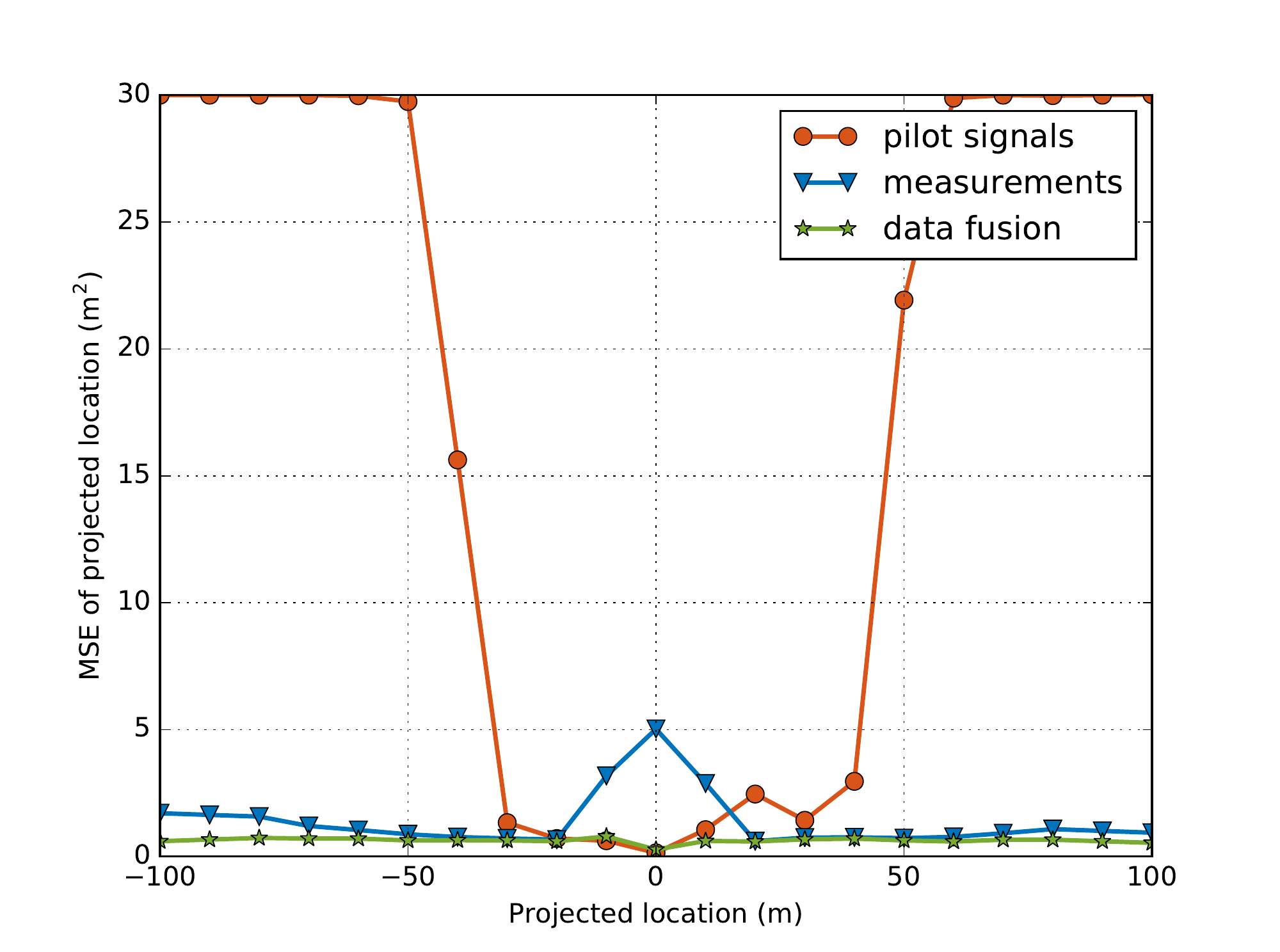}
		\label{fig:location_error_fault}
	}
	\subfigure[MSE of velocity estimation.]{		
		\includegraphics[width=3.0in]{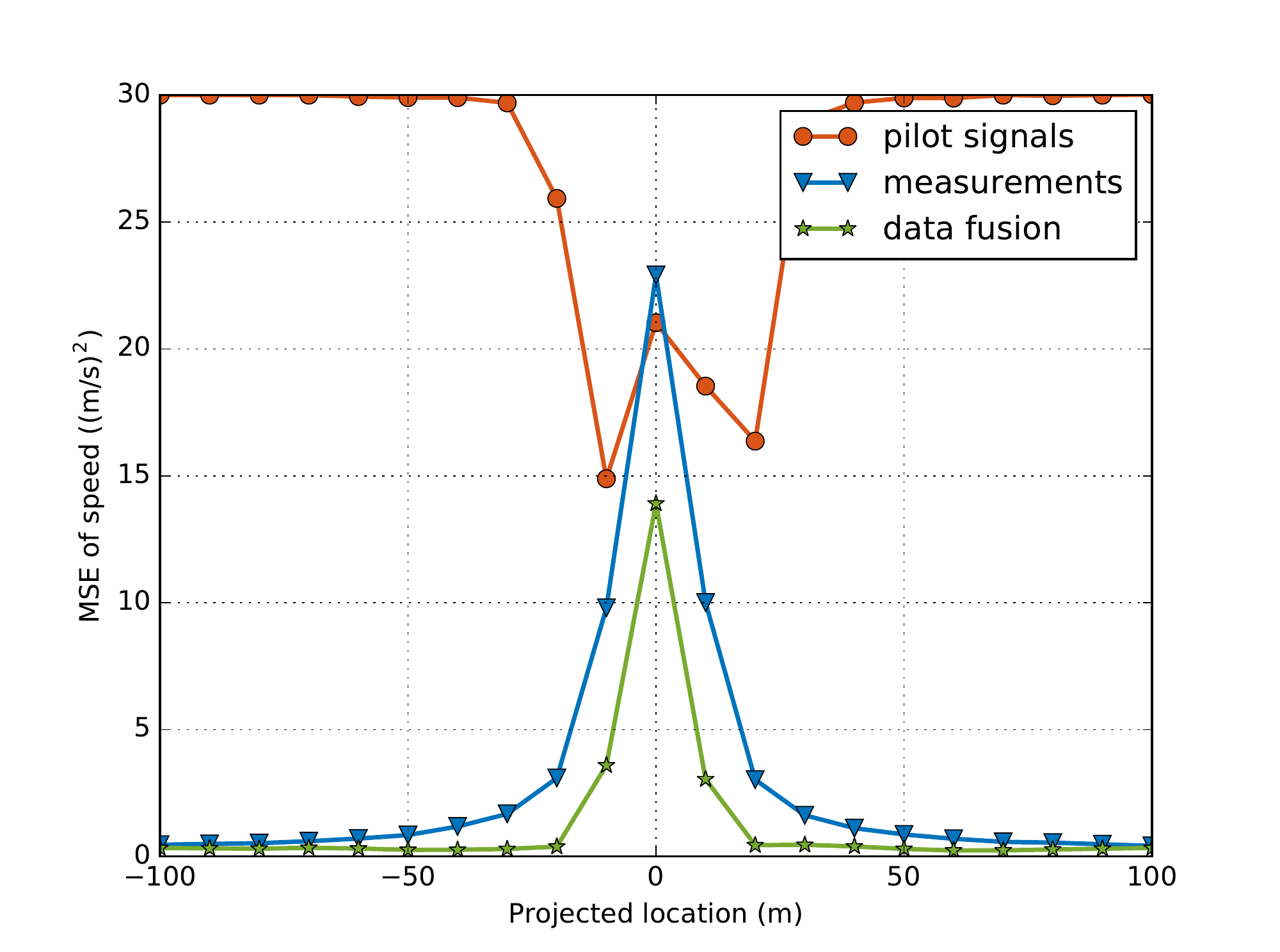}
		\label{fig:velocity_error_fault}
	}
	\caption{Non-linear tracks (linear estimations): the MSE versus projected location $ x $.}
\end{figure}

\subsection{Non-linear Mapping}

To verify the effectiveness of our proposed piece-wise function, we consider several regular regressors as comparisons, i.e., random forest (RF) with $ 100 $ decision trees, support vector machine (SVM) with linear kernels, second-order polynomial regression (Poly), and a two hidden-layer MLP where each layer has $ 512 $ neurons. The simulation result is given in Table~\ref{tab:mae}, the RF and SVM cannot perform well. The Poly is feasible, and the model complexity is very low (only three parameters). Meanwhile, the polynomial order must be known, otherwise a low-order or a high-order Poly performs badly. The MLP is also feasible, but at the cost of huge model and computation complexities. Our proposed method achieves the best trade-off between fitting precision and model/computation complexity, and does not require any priors. Thus, we claim interpretability and simplicity.

\begin{table}
	\centering 
	\setlength{\tabcolsep}{0.5mm}{
		\scriptsize
		\centering
		\caption{Fitting error measured by MAE} 
		\begin{tabular}{|c|c|c|c|c|c|} 
			\toprule
			Regressor & RF & SVM & Poly & MLP & proposed\\ 
			\midrule
			MAE & $ 3.05 $ & $ 2.29 $ & $ 0.22 $ & $ 6.44 \times 10^{-2} $ & $ 8.14 \times 10^{-3} $\\
			\bottomrule 
		\end{tabular} 
		\label{tab:mae}}
\end{table}

\begin{table}
	\centering 
	\setlength{\tabcolsep}{0.5mm}{
	\scriptsize
	\centering
	\caption{Average spectral efficiency and beam prediction accuracy} 
	\begin{tabular}{ccccc} 
		\toprule  
		& Optimal & \makecell[c]{Data\\fusion} & Measurements & \makecell[c]{Pilot\\signals}\\ 
		\midrule
		\makecell[c]{Spectral efficiency\\(bps/Hz)} & $ 13.39 $ & $ 13.37 $ & $ 13.29 $ & $ 12.66 $\\
		\makecell[c]{Beam prediction\\accuracy} & $ 99.99\% $ & $ 98.68\% $ & $ 97.64\% $ & $ 93.98\% $\\
		\bottomrule 
	\end{tabular} 
	\label{tab:sumrate_miss_rate}}
\end{table}

\subsection{Spectral Efficiency and beam prediction accuracy}

The results of spectral efficiency (SE) and beam prediction accuracy\footnote{The averaged simulation results with linear and non-linear tracks are demonstrated, since their performance are highly similar.} are listed in Table~\ref{tab:sumrate_miss_rate}. Due to the influence of acceleration, the miss-alignment of optimal occurs at a probability of $ 7.05 \times 10^{-5} $, and the optimal accuracy is $ 99.99\% $. The proposed data fusion method outperforms the methods by measurements and received pilot signals, in terms of both SE and beam prediction accuracy. We also notice that the SE and beam prediction accuracy of the proposed method are close to the optimal.

\subsection{Overhead and Throughput}

Consider MT specific downlink and uplink overheads, the overheads of BA/T and proposed beam prediction grow linearly with MT number~\cite{beam_manage}. We define overhead cost ratio as proportion of overhead occupied in time frequency resource. A real number is quantized by $ 32 $ bits. Consider a time division duplex system, the proposed beam prediction along with the baseline are demonstrated in Table~\ref{tab:overhead}. Compared to BA/T, the proposed beam prediction consumes near-zero overheads, and thus the corresponding effective throughput (reducing downlink/uplink overhead) is higher when the MT number is greater than $ 4 $. Additionally, the delay loss in BA/T is about $ 20 \; \textup{ms} $, while the proposed beam prediction has zero delay. The simulation results validate the effectiveness of our proposed scheme.

\begin{table}
	\centering 
	\setlength{\tabcolsep}{0.5mm}{
		\scriptsize
		\centering
		\caption{Overhead cost ratio and mean effective throughput} 
		\begin{tabular}{ccccccc} 
			\toprule  
			& MT number & $ 4 $ & $ 10 $ & $ 20 $ & $ 50 $ & $ 100 $\\ 
			\midrule
			\multirow{2}{*}{\makecell[c]{Overhead\\cost ratio}} & BA/T & $ 3.21\% $ & $ 8.04\% $ & $ 16.1\% $ & $ 40.2\% $ & $ 80.5\% $\\
			& Beam prediction & $ 0.0705\% $ & $ 0.183\% $ & $ 0.391\% $ & $ 1.164\% $ & $ 2.97\% $\\
			\midrule
			\multirow{2}{*}{\makecell[c]{Mean effective\\throughput (Mbps)}} & BA/T & $ 1036.30 $ & $ 397.65 $ & $ 177.49 $ & $ 49.78 $ & $ 7.97 $\\
			& Beam prediction & $ 1069.03 $ & $ 427.37 $ & $ 209.53 $ & $ 80.83 $ & $ 39.24 $\\
			\bottomrule 
		\end{tabular} 
		\label{tab:overhead}}
\end{table}

\begin{figure*}[h]
	\centering
	\includegraphics[width = 7.0in]{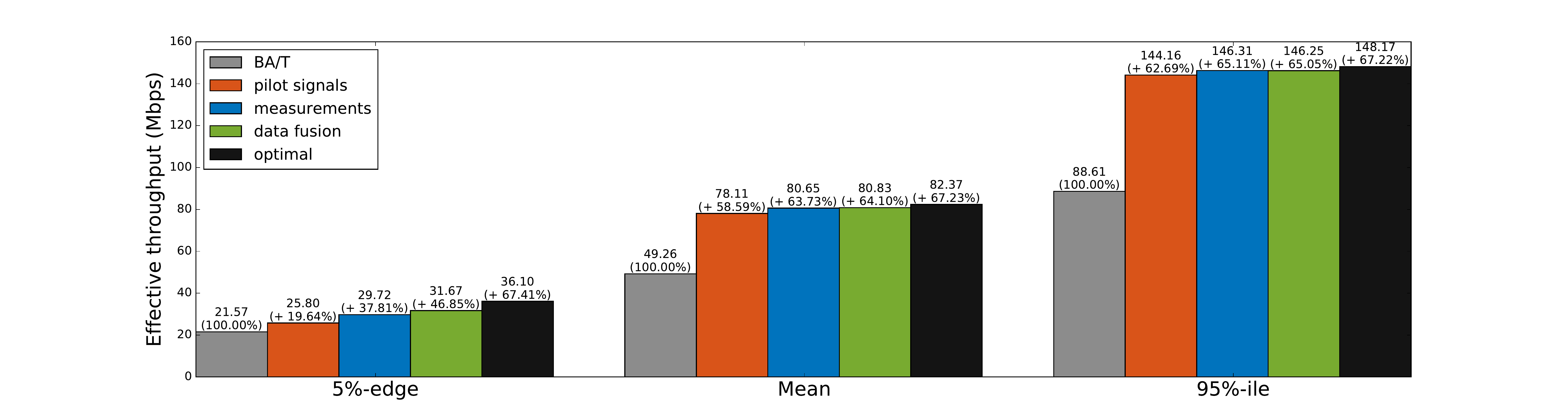}
	\caption{Effective throughput of MTs}
	\label{fig:throughput}
\end{figure*}

Furthermore, the effective throughput of $ 50 $ MTs is illustrated in Fig.~\ref{fig:throughput}. $ 5\% $-edge MT on the left side denotes the lowest $ 5\% $ MT throughput while $ 95\% $-ile denotes the highest $ 5\% $, and the middle is average throughput. Generally, the mean and $ 95\% $-ile throughputs of predictable methods are greatly improved by about $ 60\% $, compared to the baseline BA/T. We also notice that the cell-edge MTs with data fusion achieve the highest throughput, and outperforms these of pilot signals and measurements.

\section{Conclusions}\label{sec:conclusion}

In an HSR scenario, the beam prediction which transformed into a parameter estimation and a cascaded hybrid beamforming was investigated. Based on the ML criterion, the parameter estimations with received pilot signals and measurements were respectively carried out by the coordinate descent method, and a data fusion module was proposed to further improve the estimation accuracy and robustness. In hybrid beamforming, the future beam directions and channel amplitudes were predicted for hybrid beamforming. Besides, the learnable non-linear mapping module was adopted for the HSR scenarios with non-linear tracks. The simulation results showed that the proposed beam prediction scheme with learnable model-based modules outperformed the one without data fusion or the non-linear mapping, in terms of effective throughput and alignment rate.

In our future work, we will consider integrated communications and sensing, as well as learnable prior information to further improve the beam prediction performance.

\section{Acknowledgments}\label{sec:acknowledgments}

We would like to thank Dr. Hengtao He and Bo Gao for valuable discussions.

\begin{appendices}
	
\section{Analysis on the Speed and Acceleration Estimation}\label{sec:appendix_a}

\begin{figure}
	\centering
	\includegraphics[width=3.0in]{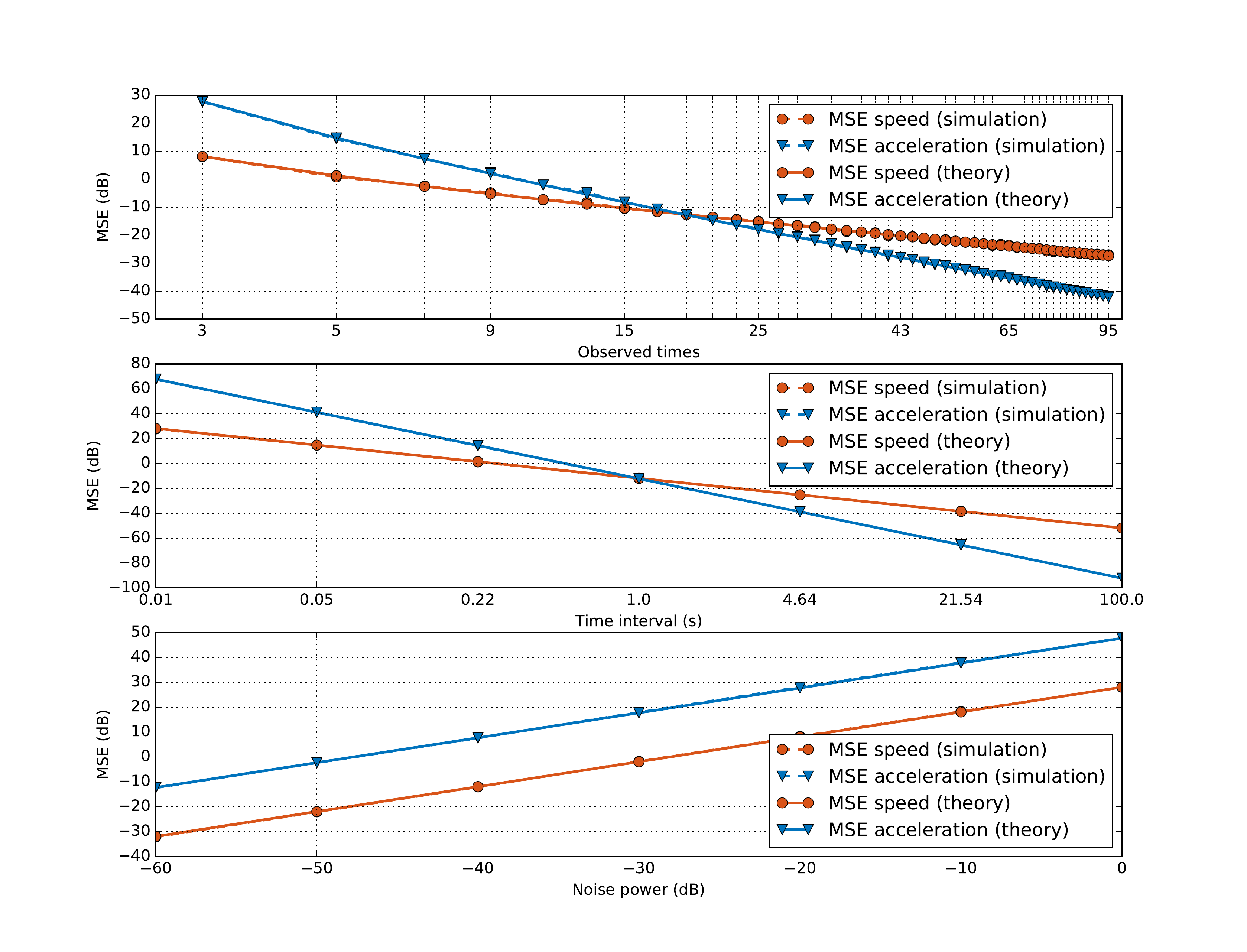}
	\caption{The MSE curves of estimation variances versus measurement times, time interval and noise power ($ L = 3 $, $ \Delta t = 0.1 \; \textup{s} $, $ \sigma_n = 0.1 $).}
	\label{fig:acceleration_analysis}
\end{figure}

Suppose that the speed and acceleration of a MT are respectively $ v $ and $ a $. Consider a linear track, according to \eqref{equ:x_l}, the projected locations are $ [x_1, \cdots, x_L]^T $ where $ L $ denotes the observed times. We have the following well-posed or over-determined equation:
\begin{align}\label{equ:over_determined_equation}
\underbrace{\begin{bmatrix}
	\Delta t & \frac{1}{2} \big(\Delta t\big)^2 \\
	\vdots & \vdots \\
	(L-1) \Delta t & \frac{1}{2} \big((L-1) \Delta t\big)^2
	\end{bmatrix}}_{\boldsymbol{A}}
\begin{bmatrix}
v \\
a
\end{bmatrix} & = 
\begin{bmatrix}
\Delta x_{n, 1} \\
\vdots \\
\Delta x_{n, L-1}
\end{bmatrix} \\
& = \begin{bmatrix}
\Delta x_1\\
\vdots \\
\Delta x_{L-1}
\end{bmatrix} +
\begin{bmatrix}
\Delta n_1 \\
\vdots \\
\Delta n_{L-1}
\end{bmatrix}
\end{align}
where $ \Delta x_i = x_{i+1} - x_1 $ and $ \Delta n_i = n_{i+1} - n_1 $, $ n_i \sim \mathcal{N}(0, \sigma_n^2), \,\forall i $, and $ \Delta t $ denotes the time interval. Multiplying $ \boldsymbol{A}^T $ at both sides of \eqref{equ:over_determined_equation}, and we have the following equation
\begin{align}\label{equ:over_determined_equation_2}
& \begin{bmatrix}
\sum_{i=1}^{L-1} (i\Delta t)^2 & \sum_{i=1}^{L-1} \frac{1}{2}(i\Delta t)^3\\
\sum_{i=1}^{L-1} \frac{1}{2}(i\Delta t)^3 & \sum_{i=1}^{L-1} \frac{1}{4}(i\Delta t)^4
\end{bmatrix}
\begin{bmatrix}
v \\
a
\end{bmatrix} \nonumber\\
= & \begin{bmatrix}
\sum_{i=1}^{L-1} (i\Delta t)\Delta x_{n, i} \\
\sum_{i=1}^{L-1} \frac{1}{2}(i\Delta t)^2\Delta x_{n, i}
\end{bmatrix}.
\end{align}
The first power, quadratic, cubic, and quartic sum formulas are expressed as
\begin{align}
S_L^1 & = \sum_{l=1}^L l = \frac{(L+1) L}{2},\label{equ:sum_1}\\
S_L^2 & = \sum_{l=1}^L l^2 = \frac{(2L+1) (L+1) L}{6},\label{equ:sum_2}\\
S_L^3 & = \sum_{l=1}^L l^3 = \frac{(L+1)^2 L^2}{4},\label{equ:sum_3}\\
S_L^4 & = \sum_{l=1}^L l^4 = \frac{(3L^2+3L-1) (2L+1) (L+1) L}{30}.\label{equ:sum_4}
\end{align}
Using \eqref{equ:sum_1}, \eqref{equ:sum_2}, \eqref{equ:sum_3}, \eqref{equ:sum_4}, and we solve the equation \eqref{equ:over_determined_equation}, obtain
\newcounter{mytempeqncnt}
\begin{figure*}[!t]
	\normalsize
	\setcounter{mytempeqncnt}{\value{equation}}
	\setcounter{equation}{57}
	\begin{align}
	v & = \frac{\Big[\sum_{i=1}^{L-1} \frac{1}{4}(i\Delta t)^4\Big] \Big[\sum_{i=1}^{L-1} (i\Delta t)\Delta x_{n, i}\Big] - \Big[\sum_{i=1}^{L-1} \frac{1}{2}(i\Delta t)^3\Big] \Big[\sum_{i=1}^{L-1} \frac{1}{2}(i\Delta t)^2\Delta x_{n, i}\Big]}{\Big[\sum_{i=1}^{L-1} (i\Delta t)^2\Big] \Big[\sum_{i=1}^{L-1} \frac{1}{4}(i\Delta t)^4\Big] - \Big[\sum_{i=1}^{L-1} \frac{1}{2}(i\Delta t)^3\Big] \Big[\sum_{i=1}^{L-1} \frac{1}{2}(i\Delta t)^3\Big]}\nonumber\\
	& = \frac{1}{\Delta t} \cdot \frac{\sum_{i=1}^{L-1} (S_{L-1}^4 i - S_{L-1}^3 i^2) \Delta x_{n, i}}{S_{L-1}^4 S_{L-1}^2 - (S_{L-1}^3)^2},\label{equ:v_derive_sim}\\
	a & = \frac{\Big[\sum_{i=1}^{L-1} (i\Delta t)^2\Big] \Big[\sum_{i=1}^{L-1} \frac{1}{2}(i\Delta t)^2\Delta x_{n, i}\Big] - \Big[\sum_{i=1}^{L-1} \frac{1}{2}(i\Delta t)^3\Big] \Big[\sum_{i=1}^{L-1} (i\Delta t)\Delta x_{n, i}\Big]}{\Big[\sum_{i=1}^{L-1} (i\Delta t)^2\Big] \Big[\sum_{i=1}^{L-1} \frac{1}{4}(i\Delta t)^4\Big] - \Big[\sum_{i=1}^{L-1} \frac{1}{2}(i\Delta t)^3\Big] \Big[\sum_{i=1}^{L-1} \frac{1}{2}(i\Delta t)^3\Big]}\nonumber\\
	& = \frac{2}{\Delta t^2} \cdot \frac{\sum_{i=1}^{L-1} (S_{L-1}^2 i^2 - S_{L-1}^3 i) \Delta x_{n, i}}{S_{L-1}^4 S_{L-1}^2 - (S_{L-1}^3)^2}.\label{equ:a_derive_sim}
	\end{align}
	\setcounter{equation}{\value{mytempeqncnt}}
	\hrulefill
	\vspace*{4pt}
\end{figure*}
\eqref{equ:v_derive_sim} and \eqref{equ:a_derive_sim}. Then, we can derive the variances of speed and acceleration as follows
\begin{align}
\sigma_v^2 = & \frac{\sigma_n^2}{\Delta t^2} \cdot \Bigg[\frac{S_{L-1}^4}{S_{L-1}^4 S_{L-1}^2 - (S_{L-1}^3)^2} \nonumber\\
& + \Big(\frac{S_{L-1}^4 S_{L-1}^1 - S_{L-1}^3 S_{L-1}^2}{S_{L-1}^4 S_{L-1}^2 - (S_{L-1}^3)^2}\Big)^2\Bigg],\tag{60}\\
\sigma_a^2 = & \frac{4 \sigma_n^2}{\Delta t^4} \cdot \Bigg[\frac{S_{L-1}^2}{S_{L-1}^4 S_{L-1}^2 - (S_{L-1}^3)^2} \nonumber\\
& + \Big(\frac{(S_{L-1}^2)^2 - S_{L-1}^3 S_{L-1}^1}{S_{L-1}^4 S_{L-1}^2 - (S_{L-1}^3)^2}\Big)^2\Bigg].\tag{61}
\end{align}
The simulation results in Fig.~\ref{fig:acceleration_analysis} is consistent with the theoretical analysis. Therefore, we have the following results:
\begin{itemize}
	\item The estimation variance of speed is proportional to the inverse of quadratic power of time interval, i.e., $ \sigma_v^2 \propto \frac{1}{\Delta t^2} $; the estimation variance of acceleration is proportional to the inverse of quartic power of time interval, i.e., $ \sigma_a^2 \propto \frac{1}{\Delta t^4} $.
	\item When the measurement times is large, the estimation variance of speed is proportional to the inverse of quadratic power of time interval, i.e., $ \lim_{L \to +\infty} \sigma_v^2 \propto \frac{1}{L^2} $; the estimation variance of acceleration is proportional to the inverse of quartic power of time interval, i.e., $ \lim_{L \to +\infty} \sigma_a^2 \propto \frac{1}{L^4} $.
	\item  Both the estimation variances of speed and acceleration are proportional to the noise power, i.e., $ \sigma_v^2 \propto \sigma_n^2, \,\, \sigma_a^2 \propto \sigma_n^2 $.
\end{itemize}
Consider a typical case where $ L = 3, \Delta t = 0.1 \; \textup{s} $, to achieve the variance of speed being $ 1 \; (\textup{m/s})^2 $, the expected absolute error of acceleration is about $ 10 $ times of that of speed. Therefore, when the period of beam prediction reaches to $ 1 \; \textup{s} $, the error is mainly caused by the error of acceleration estimation rather than that of speed, i.e., $ \sigma_a^2 \gg \sigma_v^2 $. On the other hand, to ensure the comfort of passengers, the HSR usually takes several hundred of seconds to speed up to maximum. The absolute maximal acceleration is usually at the level of $ 0.1 \; \textup{m/s}^2 $. Consider a prediction period around $ 1 \; \textup{s} $, the influence caused by acceleration is negligible. 

In summary, although the analytical model in this section is much simpler than the investigated problem, it still can prove that effect of acceleration is negligible, and the corresponding accurate estimation is infeasible, with limited observed times, time interval and prediction period.

\section{Analysis on the Expected Fitting Error}\label{sec:appendix_b}

We investigate the expected fitting error in this part, and we use $ \Theta^{\textup{opt}} $ denote the learned optimal parameter set of fitting function $ g $. Furthermore, we consider a practical track function which is modeled as an arbitrary quadratic function $ f(x) = a x^2 + b x + c $.

\subsection{Quadratic Curves}

\begin{figure}
	\centering
	\includegraphics[width=3.0in]{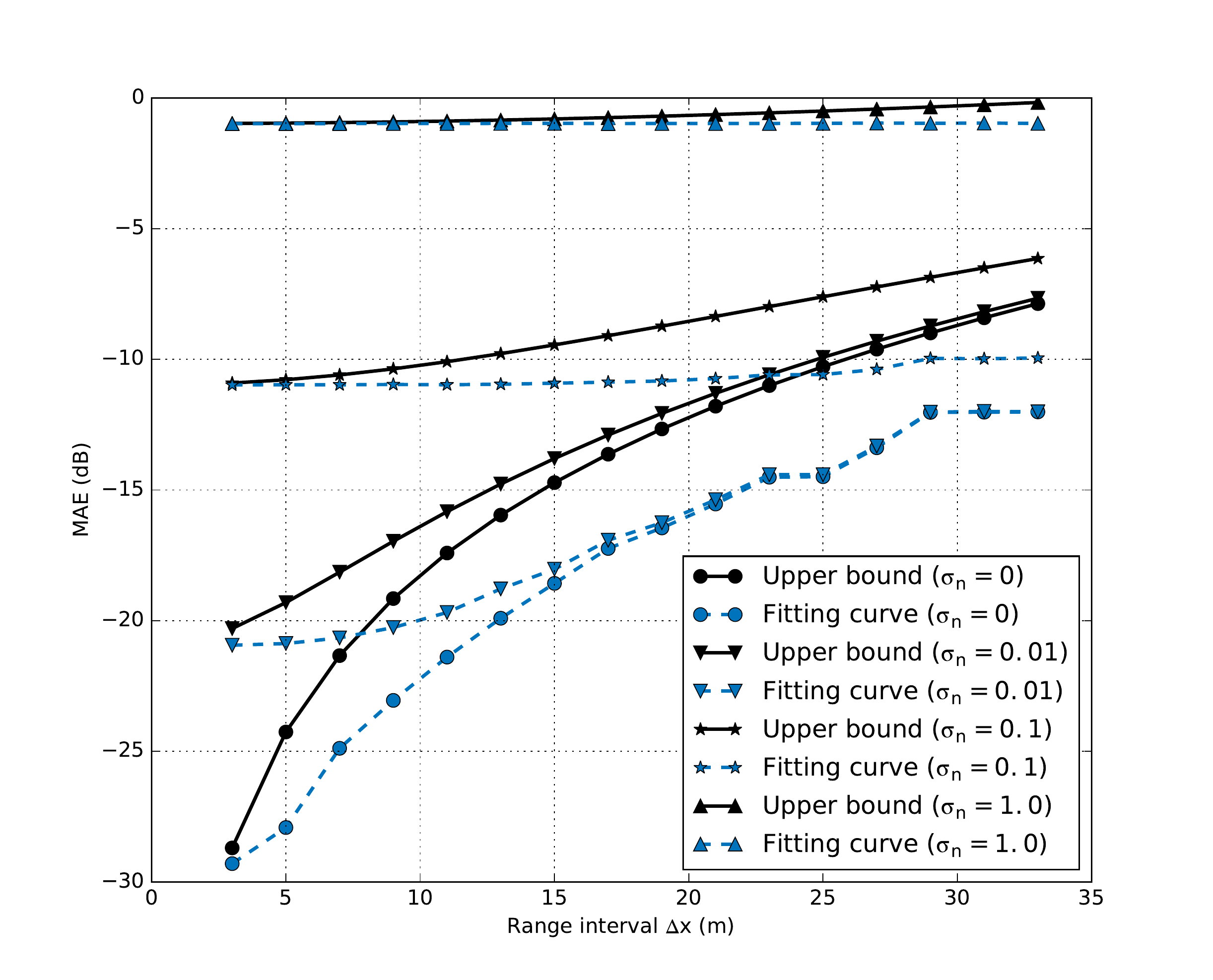}
	\caption{The fitting error versus range interval $ \Delta x $.}
	\label{fig:fitting_error}
\end{figure}

The expectation of $ e $ in \eqref{equ:loss_fitting} can be rewritten as
\begin{equation}\label{equ:inequal_1}
\begin{split}
e & = \mathbb{E}_{x \sim \mathcal{X}, n \sim \mathcal{N}}\Big\{\Big|f(x) + n - g(x; \Theta^{\textup{opt}})\Big| \Big\}\\
& = \mathbb{E}_{n \sim \mathcal{N}} \Big\{\frac{1}{2 r_{\textup{max}}} \int_{-r_{\textup{max}}}^{r_{\textup{max}}} \Big|f(x) + n - g(x; \Theta^{\textup{opt}})\Big| dx\Big\}\\
& \leq \frac{1}{2 r_{\textup{max}}} \int_{-r_{\textup{max}}}^{r_{\textup{max}}} \Big|f(x) - g(x; \Theta^{\textup{opt}})\Big| dx + \mathbb{E}_{n \sim \mathcal{N}}\Big\{|n|\Big\}\\
& = \frac{1}{2 r_{\textup{max}}} \sum_{i=1}^{N_x} \int_{x_i}^{x_i + \Delta x} \Big|f(x) - g_i(x; \Theta^{\textup{opt}})\Big| dx + \sigma_n\sqrt{\frac{2}{\pi}},\\
\end{split}\tag{62}
\end{equation}
where $ \mathcal{X} $ is a uniform distribution on the support set of $ x $. The inequality in \eqref{equ:inequal_1} is derived by trigonometric inequality. Consider a sub-optimal case where fixed start-point is $ f(x_i) $ and end-point is $ f(x_i+\Delta x) $ for the $ i $-th piece-wise function $ g_i(x; \Theta) $, the expression of $ e $ in \eqref{equ:inequal_1} can be further scaled as follows
\begin{equation}\label{equ:inequal_2}
\begin{split}
e \leq & \frac{1}{2 r_{\textup{max}}} \sum_{i=1}^{N_x} \underbrace{\int_{x_i}^{x_i + \Delta x} \Big|f(x) - \big[f(x_i) + k_i(x - x_i)\big]\Big| dx}_{e_i}\\
& + \sigma_n\sqrt{\frac{2}{\pi}}.\\
\end{split}\tag{63}
\end{equation}
We substitute the expression of $ f(x) $ into \eqref{equ:inequal_2}, and easily we have \eqref{equ:inequal_quadratic}.
\begin{figure*}[!t]
	\normalsize
	\setcounter{mytempeqncnt}{\value{equation}}
	\setcounter{equation}{63}
	\begin{equation}\label{equ:inequal_quadratic}
	e \leq \frac{1}{2 r_{\textup{max}}} \sum_{i=1}^{N_x} \int_{x_i}^{x_i + \Delta x} \Big|f(x) - \big[f(x_i) + \frac{f(x_i + \Delta x) - f(x_i)}{\Delta x}(x - x_i)\big]\Big| dx\Bigg|_{f(x) = a x^2 + b x + c} + \sigma_n\sqrt{\frac{2}{\pi}} = \frac{|a| \Delta x^2}{6} + \sigma_n\sqrt{\frac{2}{\pi}}.
	\end{equation}
	\setcounter{equation}{\value{mytempeqncnt}}
	\hrulefill
	\vspace*{4pt}
\end{figure*}
The expected loss is composed of two parts: observed noise and fitting error. Ignoring the influence of noise, the error $ e $ is in proportional to $ |a| $ and $ \Delta x^2 $. As shown in Fig.~\ref{fig:fitting_error}, generally the fitting error gap between the theoretical curve and the simulation curve is about $ 3 $ -- $ 4 \; \textup{dB} $. When the range interval $ \Delta x $ is decreased to $ 3 \; \textup{m} $, this gap is sharply reduced. The updated parameter point during training is oscillating around the optimal, thus it is difficult to obtain the fitting curve very close to the track curve.

\end{appendices}

\newpage
\bibliographystyle{IEEEtran}
\bibliography{References}

\begin{thebibliography}{10}
\providecommand{\url}[1]{#1}
\csname url@samestyle\endcsname
\providecommand{\newblock}{\relax}
\providecommand{\bibinfo}[2]{#2}
\providecommand{\BIBentrySTDinterwordspacing}{\spaceskip=0pt\relax}
\providecommand{\BIBentryALTinterwordstretchfactor}{4}
\providecommand{\BIBentryALTinterwordspacing}{\spaceskip=\fontdimen2\font plus
\BIBentryALTinterwordstretchfactor\fontdimen3\font minus
  \fontdimen4\font\relax}
\providecommand{\BIBforeignlanguage}[2]{{%
\expandafter\ifx\csname l@#1\endcsname\relax
\typeout{** WARNING: IEEEtran.bst: No hyphenation pattern has been}%
\typeout{** loaded for the language `#1'. Using the pattern for}%
\typeout{** the default language instead.}%
\else
\language=\csname l@#1\endcsname
\fi
#2}}
\providecommand{\BIBdecl}{\relax}
\BIBdecl

\bibitem{GlobeCom21}
{Y. Zhang, S. Liu, Z. Lu, F. Meng, and Y. Huang}, ``Learning-aided beam
  management for mmwave high-speed railway networks,'' in \emph{Proc. 40-th
  IEEE Global Commun. Conf. (GLOBECOM'21): Signal Process. Commun. Symp.,
  \textup{Madrid, Spain}}, Dec. 2021, pp. 1--6.

\bibitem{5490974}
A.~Ghosh, R.~Ratasuk, B.~Mondal, N.~Mangalvedhe, and T.~Thomas,
  ``{LTE-advanced: Next-generation wireless broadband technology [Invited
  Paper]},'' \emph{IEEE Wireless Commun.}, vol.~17, no.~3, pp. 10--22, 2010.

\bibitem{7959169}
M.~Xiao, S.~Mumtaz, Y.~Huang, L.~Dai, Y.~Li, M.~Matthaiou, G.~K. Karagiannidis,
  E.~Björnson, K.~Yang, C.-L. I, and A.~Ghosh, ``Millimeter wave
  communications for future mobile networks,'' \emph{IEEE J. Sel. Areas
  Commun.}, vol.~35, no.~9, pp. 1909--1935, 2017.

\bibitem{Goodfellow2016Deep}
I.~Goodfellow, Y.~Bengio, and A.~Courville, \emph{Deep Learning}.\hskip 1em
  plus 0.5em minus 0.4em\relax The MIT Press, 2016.

\bibitem{8054694}
T.~O’Shea and J.~Hoydis, ``An introduction to deep learning for the physical
  layer,'' \emph{IEEE Trans. Trans. Cogn. Commun. Netw.}, vol.~3, no.~4, pp.
  563--575, 2017.

\bibitem{9120241}
F.~Meng, P.~Chen, L.~Wu, and J.~Cheng, ``Power allocation in multi-user
  cellular networks: {Deep} reinforcement learning approaches,'' \emph{IEEE
  Trans. Wireless Commun.}, vol.~19, no.~10, pp. 6255--6267, 2020.

\bibitem{8353153}
H.~He, C.-K. Wen, S.~Jin, and G.~Y. Li, ``Deep learning-based channel
  estimation for beamspace {mmWave} massive {MIMO} systems,'' \emph{IEEE
  Wireless Commun. Lett.}, vol.~7, no.~5, pp. 852--855, 2018.

\bibitem{9018199}
------, ``Model-driven deep learning for {MIMO} detection,'' \emph{IEEE Trans.
  Signal Process.}, vol.~68, pp. 1702--1715, 2020.

\bibitem{8742579}
A.~Zappone, M.~Di~Renzo, and M.~Debbah, ``Wireless networks design in the era
  of deep learning: {Model-based}, {AI}-based, or both?'' \emph{IEEE Trans.
  Commun.}, vol.~67, no.~10, pp. 7331--7376, 2019.

\bibitem{8715338}
H.~He, S.~Jin, C.-K. Wen, F.~Gao, G.~Y. Li, and Z.~Xu, ``Model-driven deep
  learning for physical layer communications,'' \emph{IEEE Wireless Commun.},
  vol.~26, no.~5, pp. 77--83, 2019.

\bibitem{6717211}
O.~E. Ayach, S.~Rajagopal, S.~Abu-Surra, Z.~Pi, and R.~W. Heath, ``Spatially
  sparse precoding in millimeter wave {MIMO} systems,'' \emph{IEEE Trans.
  Wireless Commun.}, vol.~13, no.~3, pp. 1499--1513, 2014.

\bibitem{7390019}
Z.~Marzi, D.~Ramasamy, and U.~Madhow, ``Compressive channel estimation and
  tracking for large arrays in {mm-Wave} picocells,'' \emph{IEEE J. Sel. Topics
  Signal Process.}, vol.~10, no.~3, pp. 514--527, 2016.

\bibitem{2017Location}
Z.~Cheng, Z.~Jing, Y.~Huang, and L.~Yang, ``Location-aided channel tracking and
  downlink transmission for {HST} massive {MIMO} systems,'' \emph{{IET
  Commun.}}, vol.~11, no.~13, pp. 2082--2088, 2017.

\bibitem{8823977}
K.~{Satyanarayana}, M.~{El-Hajjar}, A.~A.~M. {Mourad}, and L.~{Hanzo}, ``Deep
  learning aided fingerprint-based beam alignment for {mmWave} vehicular
  communication,'' \emph{IEEE Trans. Veh. Technol.}, vol.~68, no.~11, pp.
  10\,858--10\,871, 2019.

\bibitem{9013296}
Y.~{Heng} and J.~G. {Andrews}, ``Machine learning-assisted beam alignment for
  {mmWave} systems,'' in \emph{2019 IEEE Global Commun. Conf. (GLOBECOM),
  \textup{Waikoloa, HI, USA}}, 2019.

\bibitem{9129762}
W.~{Xu}, F.~{Gao}, S.~{Jin}, and A.~{Alkhateeb}, ``{3D} scene-based beam
  selection for {mmWave} communications,'' \emph{IEEE Wireless Commun. Lett.},
  vol.~9, no.~11, pp. 1850--1854, 2020.

\bibitem{jonathan2019deep}
N.~M. Jonathan, W.~Yuyang, G.-P. Nuria, and J.~R.~H. W., ``Deep learning-based
  beam alignment in {mmWave} vehicular networks,'' \emph{Proc. IEEE Int.
  Acoust., Speech, Signal Process. (ICASSP), \textup{Brighton, UK}}, pp.
  8569--8573, 2019.

\bibitem{Sutton}
R.~Sutton and A.~Barto, \emph{Reinforcement Learning: An Introduction}.\hskip
  1em plus 0.5em minus 0.4em\relax MIT Press, 2018.

\bibitem{8842625}
W.~Wu, N.~Cheng, N.~Zhang, P.~Yang, W.~Zhuang, and X.~Shen, ``Fast mmwave beam
  alignment via correlated bandit learning,'' \emph{IEEE Trans. Wireless
  Commun.}, vol.~18, no.~12, pp. 5894--5908, 2019.

\bibitem{9069211}
J.~Zhang, Y.~Huang, Y.~Zhou, and X.~You, ``Beam alignment and tracking for
  millimeter wave communications via bandit learning,'' \emph{IEEE Trans.
  Commun.}, vol.~68, no.~9, pp. 5519--5533, 2020.

\bibitem{9269463}
J.~{Zhang}, Y.~{Huang}, J.~{Wang}, X.~{You}, and C.~{Masouros}, ``Intelligent
  interactive beam training for millimeter wave communications,'' \emph{IEEE
  Trans. Wireless Commun.}, pp. 1--1, 2020.

\bibitem{8410591}
J.~{Ma}, S.~{Zhang}, H.~{Li}, F.~{Gao}, and S.~{Jin}, ``Sparse bayesian
  learning for the time-varying massive {MIMO} channels: {Acquisition} and
  tracking,'' \emph{IEEE Trans. Commun.}, vol.~67, no.~3, pp. 1925--1938, 2019.

\bibitem{8984259}
L.~{Yan}, X.~{Fang}, L.~{Hao}, and Y.~{Fang}, ``A fast beam alignment scheme
  for dual-band {HSR} wireless networks,'' \emph{IEEE Trans. Veh. Technol.},
  vol.~69, no.~4, pp. 3968--3979, 2020.

\bibitem{7295467}
B.~Ai, K.~Guan, M.~Rupp, T.~Kurner, X.~Cheng, X.-F. Yin, Q.~Wang, G.-Y. Ma,
  Y.~Li, L.~Xiong, and J.-W. Ding, ``Future railway services-oriented mobile
  communications network,'' \emph{IEEE Commun. Mag.}, vol.~53, no.~10, pp.
  78--85, 2015.

\bibitem{7811845}
H.~Song, X.~Fang, and Y.~Fang, ``Millimeter-wave network architectures for
  future high-speed railway communications: {Challenges} and solutions,''
  \emph{IEEE Wireless Commun.}, vol.~23, no.~6, pp. 114--122, 2016.

\bibitem{8458146}
M.~Giordani, M.~Polese, A.~Roy, D.~Castor, and M.~Zorzi, ``A tutorial on beam
  management for {3GPP NR} at {mmWave} frequencies,'' \emph{IEEE Commun.
  Surveys Tuts.}, vol.~21, no.~1, pp. 173--196, 2019.

\bibitem{beam_manage}
{\textit{Enhancements on predictable mobility for beam management}},
  \emph{\textup{3GPP TSG RAN WG 90 e-Meeting. RP-202675, ZTE}}, Dec. 2020.

\bibitem{38901}
3GPP, ``{Study on channel model for frequencies from 0.5 to 100 GHz},''
  \emph{\textup{3GPP TR 38.901}}, Jan. 2020, version 16.1.0.

\bibitem{2012Sparsity}
A.~Beck and Y.~C. Eldar, ``Sparsity constrained nonlinear optimization:
  {Optimality} conditions and algorithms,'' \emph{SIAM J. Optim.}, vol.~23,
  no.~3, pp. 1480--1509, 2012.

\bibitem{attention}
A.~Vaswani, N.~Shazeer, N.~Parmar, J.~Uszkoreit, L.~Jones, A.~N. Gomez,
  u.~Kaiser, and I.~Polosukhin, ``Attention is all you need,'' in \emph{Proc.
  Int. Conf. Neural Inf. Process. Syst. (NIPS)}, 2017, p. 6000–6010.

\bibitem{beam_enhance}
{\textit{Moderator summary for multi-beam enhancement: EVM}},
  \emph{\textup{3GPP TSG RAN WG1 102 e-Meeting. R1-2007151, Moderator
  (Samsung)}}, Aug. 2020.

\bibitem{Ericsson}
A.~{Zaidi}, R.~{Baldemair}, M.~{Andersson}, S.~{Faxér}, V.~{Molés-Cases}, and
  Z.~{Wang}, ``{5G} new radio: {Designing} for the future,'' \emph{Ericsson
  Technol. Rev.}, 2017.

\end{thebibliography}

\begin{IEEEbiography}[{\includegraphics[width=1in,height=1.25in,clip,keepaspectratio]{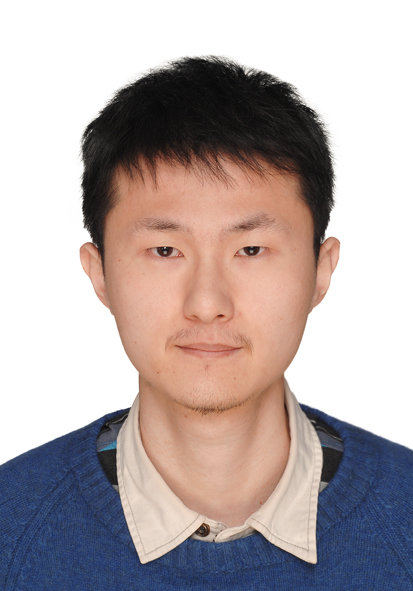}}]{Fan Meng}
	received the B.S. degree in 2015 from the school of electronic engineering in the University of Electronic Science and Technology of China, UESTC, and the Ph.D. degree in 2020 from the school of Information Science and Engineering, Southeast University, China. He is now a wireless communication researcher in Purple Mountain Laboratories. His main research topic is applying machine learning techniques in the wireless communication systems. Further research interests include machine learning in general, joint demodulation and equalization, resource allocation, intelligent beamforming and precoding.	
\end{IEEEbiography}

\begin{IEEEbiography}[{\includegraphics[width=1in,height=1.25in,clip,keepaspectratio]{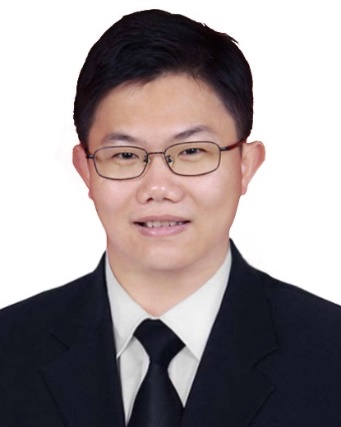}}]{Shengheng Liu (S'14-M'17)}
	received the B.Eng. and Ph.D. degrees in electronics engineering from the School of Information and Electronics, Beijing Institute of Technology, Beijing, China, in 2010 and 2017, respectively.
	
	Dr. Liu is currently an Associate Professor with the School of Information Science and Engineering, Southeast University (SEU), Nanjing, China. Prior to joining SEU, he held a postdoctoral position at the Institute for Digital Communications, The University of Edinburgh, Edinburgh, U.K., from 2017 to 2018. He also worked as a Visiting Research Associate from 2015 to 2016 at the Department of Electrical and Computer Engineering, Temple University, Philadelphia, PA, USA, under the support of the China Scholarship Council. Dr. Liu is a recipient of the 2017 National Excellent Doctoral Dissertation Award from the China Institute of Communications. His research interests mainly focus on intelligent sensing and wireless communications.
\end{IEEEbiography}

\begin{IEEEbiography}[{\includegraphics[width=1in,height=1.25in,clip,keepaspectratio]{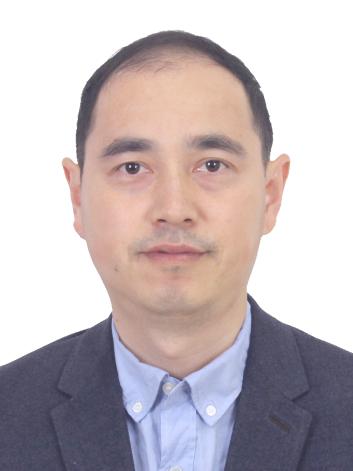}}]{Yongming Huang (M'10-SM'16)}
	received the B.S. and M.S. degrees from Nanjing University, Nanjing, China, in 2000 and 2003, respectively, and the Ph.D. degree in electrical engineering from Southeast University, Nanjing, in 2007.
	
	Since March 2007 he has been a faculty in the School of Information Science and Engineering, Southeast University, China, where he is currently a full professor. He has also been the Director of the Pervasive Communication Research Center, Purple Mountain Laboratories, since 2019. From 2008 to 2009, he visited the Signal Processing Lab, Royal Institute of Technology (KTH), Stockholm, Sweden. He has published over 200 peer-reviewed papers, hold over 80 invention patents. His current research interests include intelligent 5G/6G mobile communications and millimeter wave wireless communications. He submitted around 20 technical contributions to IEEE standards, and was awarded a certificate of appreciation for outstanding contribution to the development of IEEE standard 802.11aj. He served as an Associate Editor for the IEEE Transactions on Signal Processing and a Guest Editor for the IEEE Journal Selected Areas in Communications. He is currently an Editor-at-Large for the IEEE Open Journal of the Communications Society and an Associate Editor for the IEEE Wireless Communications Letters.
\end{IEEEbiography}

\begin{IEEEbiography}[{\includegraphics[width=1in,height=1.25in,clip,keepaspectratio]{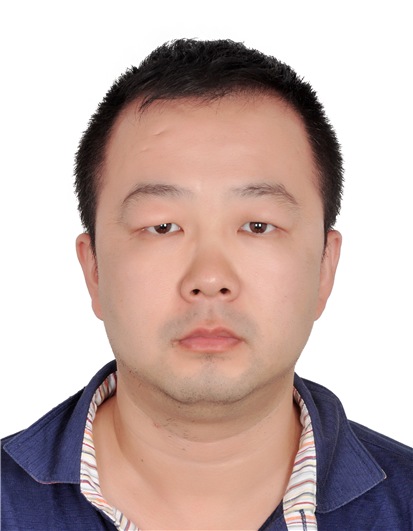}}]{Zhaohua Lu}
	(lu.zhaohua@zte.com.cn), Ph.D., graduated from Tianjin University in 2006. He is senior wireless communication system research expert of ZTE Corporation and has long been engaged in the field of wireless communication system design and the key technologies of the physical layer. He has many technical contributions, papers, and patents in interference mitigation in the MIMO field.
\end{IEEEbiography}

\end{document}